\documentclass[fleqn,usenatbib]{mnras}

\usepackage[T1]{fontenc}

\DeclareRobustCommand{\VAN}[3]{#2}
\let\VANthebibliography\thebibliography
\def\thebibliography{\DeclareRobustCommand{\VAN}[3]{##3}\VANthebibliography}

\usepackage{graphicx}	% Including figure files
\usepackage{amsmath}	% Advanced maths commands
\usepackage{amssymb}	% Extra maths symbols

\usepackage{epsfig}
\usepackage{aas_macros}
\usepackage{natbib}
\usepackage{times}
\usepackage{graphicx, bm, xcolor}
\usepackage{verbatim}
\usepackage[all]{hypcap}
\usepackage{xspace}
\usepackage{multirow}
\usepackage{epstopdf}
\usepackage{pdflscape}
\usepackage{xspace}

\usepackage{newtxtext,newtxmath}
\newcommand{\msun}{\mbox{M$_\odot$}}

\newcommand{\kpc}{\mbox{${\rm kpc}$}}

\newcommand{\feh}{\mbox{$[{\rm Fe}/{\rm H}]$}}
\newcommand{\be}{\begin{equation}}
\newcommand{\ee}{\end{equation}}
\newcommand{\bea}{\begin{eqnarray}}
\newcommand{\eea}{\end{eqnarray}}

\newcommand*\code[1]{{\fontfamily{lmtt}\selectfont #1}}

\newcommand{\emosaics}{{\sc E-MOSAICS}\xspace}

\newcommand{\eagle}{{\sc EAGLE}\xspace}
\newcommand{\illustris}{{\sc Illustris}\xspace}

\newcommand{\illustrisTNG}{{\sc Illustris-TNG}\xspace}

\newcommand{\subfind}{\code{SUBFIND}\xspace}
\newcommand{\fof}{\code{FoF}\xspace}

\title[GC spatial distributions in E-MOSAICS]{Globular clusters as tracers of the dark matter halo: insights from the E-MOSAICS simulations}
\author[M. Reina-Campos et al.]{Marta~Reina-Campos$^{1,2,3}$\thanks{reinacampos@mcmaster.ca}, Sebastian~Trujillo-Gomez$^{3}$, Alis~J.~Deason$^{4,5}$,  J.~M.~Diederik~Kruijssen$^{3}$,  \newauthor
Joel~L.~Pfeffer$^{6}$, Robert~A.~Crain$^{7}$, Nate~Bastian$^{8,9,7}$, and Meghan~E.~Hughes$^{7}$\\
$^{1}$Department of Physics \& Astronomy, McMaster University, 1280 Main Street West, Hamilton, L8S 4M1, Canada\\
$^{2}$Canadian Institute for Theoretical Astrophysics (CITA), University of Toronto, 60 St George St, Toronto, M5S 3H8, Canada\\
$^{3}$Astronomisches Rechen-Institut, Zentrum f\"{u}r Astronomie der Universit\"{a}t Heidelberg, M\"{o}nchhofstra\ss e 12-14, 69120 Heidelberg, Germany\\
$^{4}$Institute for Computational Cosmology, Department of Physics, University of Durham, South Road, Durham DH1 3LE, UK\\
$^{5}$Centre for Extragalactic Astronomy, Department of Physics, University of Durham, South Road, Durham DH1 3LE, UK\\
$^{6}$International Centre for Radio Astronomy Research (ICRAR), M468, University of Western Australia, 35 Stirling Hwy, Crawley, WA 6009, Australia\\
$^{7}$Astrophysics Research Institute, Liverpool John Moores University, 146 Brownlow Hill, Liverpool L3 5RF, UK\\
$^{8}$Donostia International Physics Center (DIPC), Paseo Manuel de Lardizabal, 4, 20018, Donostia-San Sebastián, Guipuzkoa, Spain\\
$^{9}$IKERBASQUE, Basque Foundation for Science, 48013, Bilbao, Spain}

\date{Accepted XXX. Received YYY; in original form ZZZ}

\pubyear{2021}

\begin{document}
\label{firstpage}
\pagerange{\pageref{firstpage}--\pageref{lastpage}}
\maketitle

\begin{abstract}
Globular clusters (GCs) are bright objects that span a wide range of galactocentric distances, and are thus probes of the structure of dark matter (DM) haloes. In this work, we explore whether the projected radial profiles of GCs can be used to infer the structural properties of their host DM haloes. We use the simulated GC populations in a sample of 166 central galaxies from the $(34.4~\rm cMpc)^3$ periodic volume of the \emosaics project. We find that more massive galaxies host stellar and GC populations with shallower density profiles that are more radially extended. In addition, the metal-poor GC subpopulations tend to have shallower and more extended profiles than the metal-rich subsamples, which we relate to the preferentially accreted origin of the metal-poor GCs. We find strong correlations between the slopes and effective radii of the radial profiles of the GC populations and the structural properties of the DM haloes, such as their power-law slopes, scale radii, and concentration parameters. Accounting for a dependence on the galaxy stellar mass decreases the scatter of the two-dimensional relations. This suggests that the projected number counts of GCs, combined with their galaxy mass, trace the density profile of the DM halo of their host galaxy. When applied to extragalactic GC systems, we recover the scale radii and the extent of the DM haloes of a sample of ETGs with uncertainties smaller than $0.2~\rm dex$. Thus, extragalactic GC systems provide a novel avenue to explore the structure of DM haloes beyond the Local Group.
% word limit: 250/250
\end{abstract}

\begin{keywords}
galaxies: star clusters: general --- globular clusters: general --- stars: formation --- galaxies: evolution --- galaxies: formation
\end{keywords}

%%%%%%%%%%%%%%%%%%%%%%%%%%%%%%%%%%%%%%%%%%%%%%%%%%

%%%%%%%%%%%%%%%%% BODY OF PAPER %%%%%%%%%%%%%%%%%%
\section{Introduction} \label{sec:intro}

Galaxies reside at the center of extended haloes of dark matter (DM) that cannot be directly traced using star light. The presence and properties of such haloes must be inferred from their gravitational influence on other objects. In order to do so, the outskirts of galaxies are the best environments. Firstly, the influence of the baryonic physics, i.e.~gas cooling, star formation and feedback, modifies the shape of the DM halo in the centres of galaxies \citep[e.g.][]{duffy10,schaller15,prada19}. Additionally, most of the mass in the halo lies at large distances from the center, suggesting that probing beyond the extent of the galaxy ($\gtrsim0.1\times r_{200}$) is required to trace the structure of the DM halo.

Diffuse stellar haloes surround galaxies and can extend up to several hundred kpc. These extended stellar populations have been found to grow mostly via the accretion of satellite galaxies \citep[e.g.][]{bullock05,abadi06,cooper10,font20}, with a few massive accretion events dominating the mass assembly \citep{deason16, monachesi19}. Although the inner regions of stellar haloes tend to be dominated by in-situ stars \citep{font11}, their (mainly) accreted origin in the outskirts and their large extent suggest that stellar haloes can be used as tracers of the DM halo of their host galaxy. \citet{pillepich14} show that the radial profiles of the stellar haloes in the \illustris simulations correlate with the profiles of their DM haloes, with more massive haloes showing shallower distributions of DM and stars \citep[see also][]{pillepich18}. The authors relate this trend to the amount of accreted mass in the galaxies, i.e.~the growth of massive galaxies is mostly linked to the accretion of satellites that can deposit their stars further out, whereas low mass galaxies grow mostly due to in-situ star formation \citep[e.g.][]{rodriguez-gomez16,qu17,clauwens18,behroozi19,davison20}. Despite this promising relation, measuring the profiles of diffuse stellar haloes out to large galactocentric radii is observationally challenging, and so this approach has a limited applicability. 

Satellite galaxies and bright globular clusters (GCs) are also tracers of the structure of DM haloes. These tracers have been used to probe the galactic outskirts via dynamical models \citep[e.g.][]{tortora16,alabi16,alabi17,poci17}. These objects are much brighter than the diffuse stellar halo, and so they can be observed out to much greater galactocentric distances.
However, these tracers are not observed in equal numbers, as central galaxies generally host fewer satellite galaxies than bright GCs. In galaxies of mass similar to the Milky Way, up to $\sim10$--$20$ satellite galaxies with luminosities $M_{r} \leq -12.3$ are observed \citep{geha17,mao21}, whereas the GC populations typically comprise $\sim200$ objects \citep[e.g.][]{peng08}. This suggests that GC populations can be ideal probes of the matter distribution at large galactocentric distances. Additionally, dynamical models require accurate tracer kinematics, and this limits the number of galaxies to which these models can be applied as spectroscopy can become challenging\footnote{Using a suite of 25 Milky Way-mass cosmological zoom-in simulations from the \emosaics project, \citet{hughes21a} find that good kinematic information of at least $150$ GCs per galaxy is required to recover the mass and radial distribution of the DM halo using dynamical models in extragalactic systems.}. In contrast, if the \textit{spatial distributions} of these tracers alone could yield information on the structure of their host DM halo, then these properties could be inferred for a much larger number of galaxies. 

A surprising result from the last couple of decades has been the strong correlations between properties of the overall GC populations and their host DM haloes. The most prominent example is the observed tight relation between the total mass in GCs and the mass of the DM halo \citep[e.g.][]{blakeslee97,peng08,spitler09,georgiev10,hudson14,harris15,harris17c}, which has been linked to the hierarchical assembly of galaxies \citep[e.g.][]{kruijssen15b,choksi18,elbadry19,bastian20}. It has also been observed that the spatial extent of extragalactic GC systems strongly increases with the effective radius of the galaxy \citep[e.g.][]{rhode07,kartha14,kartha16} and with the extent of the DM halo \citep[e.g.][]{forbes17,hudson18}, implying that more massive galaxies host more extended GC populations. Given that the fraction of accreted mass also increases with mass towards massive galaxies \citep[e.g.][]{rodriguez-gomez16,qu17,behroozi19}, this suggests that GCs can be used as tracers of the detailed structural properties of the DM haloes of their host galaxies.

Based on these observational results, in this work we study the spatial distribution of stellar clusters and field stars around central galaxies, and explore how the GC spatial distributions trace the DM halo of their host galaxies. For this, we use the simulated populations of stellar clusters and their host galaxies from the $(34.4~{\rm cMpc})^3$ periodic volume of the \emosaics project \citep[][Crain et al. in prep.]{pfeffer18,kruijssen19a}. These simulations self-consistently model the formation and evolution of stellar cluster populations alongside their host galaxies in a cosmological context, and so they naturally provide the spatial information of the GCs and their host DM haloes across a broad range of galaxy masses and environments. 

We describe the simulation setup in Section~\ref{sec:emosaics}, and discuss the spatial distributions of stars and GCs in Section~\ref{sec:spatdistr}. In Section~\ref{sec:fit-profiles} we characterize the radial profiles of stars, GCs and DM, and explore possible correlations with the structural properties of DM haloes in Section~\ref{sec:correlations}. The findings of this work are summarized in Section~\ref{sec:conclusions}.

\section{The E-MOSAICS project}\label{sec:emosaics}

\subsection{The E-MOSAICS model}

The \emosaics project \citep[MOdelling Star cluster population Assembly In Cosmological Simulations within \eagle,][]{pfeffer18,kruijssen19a} combines a subgrid description of the formation and evolution of stellar clusters \citep[][]{kruijssen11,pfeffer18} with the state-of-the-art \eagle galaxy formation model \citep[Evolution and Assembly of GaLaxies and their Environments,][]{schaye15,crain15}. By modelling stellar clusters and their host galaxies simultaneously, these simulations allow us to study their formation and assembly across cosmic history. This model has been found to reproduce many properties of both the young and old cluster populations observed in the local Universe \citep[e.g.][]{kruijssen19a,pfeffer19b,hughes20} and has led to several predictions for the conditions leading to the formation of GCs at high redshift \citep[e.g.][]{pfeffer19a,reina-campos19,keller20b}. Additionally, the model has allowed the use of GC populations to trace the formation and assembly history of their host galaxy \citep[e.g.][]{kruijssen19a,hughes19,pfeffer20,trujillo-gomez21}. We have recently applied these insights to the GC population of the Milky Way, resulting in the quantitative reconstruction of its merger tree \citep[][]{kruijssen19b,kruijssen20}.

In the \emosaics simulations, a stellar cluster population can form within every newborn star particle in a subgrid fashion. The formation of stellar clusters belonging to such a population is described in terms of two ingredients: the fraction of star formation in bound clusters \citep[i.e.~the cluster formation efficiency or CFE,][]{bastian08,kruijssen12d}, and the upper truncation mass scale of the \citet{schechter76} initial cluster mass function \citep{reina-campos17}. These models have been shown to accurately describe cluster formation in the local Universe \citep[e.g.][]{adamo15b,reina-campos17,messa18,adamo20}. These ingredients define the stellar mass budget to form clusters from, and the shape of the initial cluster mass function, respectively, and are influenced by the local natal gas conditions of newborn stars, such that higher gas pressure environments lead to the formation of a larger number of clusters with larger masses. After their formation, stellar clusters evolve due to stellar evolution, two-body interactions and tidal shocks. Finally, the complete disruptive effects of dynamical friction are applied in post-processing. For more details on the models, we refer the reader to \citet{pfeffer18} and \citet{kruijssen19a}.

\subsection{Selecting the galaxy sample}\label{subsec:select-data}

We study all central galaxies with stellar masses $M_{\star}\geq10^8~\msun$ within the periodic cosmological volume of $(34.4~{\rm cMpc})^3$ from the \emosaics project (Crain et al. in prep.; see first results in \citealt{bastian20} and \citealt{hughes21b}). For details on the numerical resolution of the simulation, we refer the reader to section 2.2.~of \citet{bastian20}. This stellar mass selection leaves us with $N=994$ galaxies. DM haloes are first identified using the \fof algorithm \citep[Friends-of-Friends,][]{davis85}, with a linking length of $0.2$ times the mean particle separation. Then, gas and stellar particles are associated to the nearest DM particle, and within each halo, the \subfind algorithm \citep{springel01b,dolag09} identifies gravitationally bound substructures. The central galaxies used in this study correspond to the most massive bound structure within each DM halo. We select both the bound and unbound DM particles within the radius $r_{200}$\footnote{It is common to describe the size of DM haloes based on the overdensity relative to the critical density enclosed within that region, $\bar \rho(r\leq r_{\rm X})/\rho_{\rm cr} = X$. In this work, we use $X = 200$ to define the halo masses and sizes,$M_{200}$ and $r_{200}$, and $X = 18\pi^2 + 82(\Omega_{\rm m}(z)-1) - 39(\Omega_{\rm m}(z)-1)^2$ to define the virial radius of the halo, $r_{\rm vir}$. The \eagle galaxy formation model adopts the cosmological parameters $\Omega_{\rm m} = 0.307$, $\Omega_{\rm b} = 0.048$, $\Omega_{\rm \Lambda} = 0.693$, and $\sigma_{8} = 0.829$ as provided by the Planck satellite (\citealt{planck14}, see also table B1 in \citealt{schaye15}).} to define the DM haloes considered.

With the aim of comparing to observations, we select stars and stellar clusters within the radial range $[1,15] \times r_{1/2 M_{\star}}$\footnote{Observations typically extend up to $\sim5$--$20$ times the stellar effective radius depending slightly on the mass of the galaxy \citep[see fig.~1 by][]{alabi16}. For simplicity, we decide to use a fixed radial range across our galaxy sample, and we explore the influence of the radial range in Appendix~\ref{app:diff-radial-range}.}, where $r_{1/2 M_{\star}}$ is the 3D stellar half-mass radius of the host galaxy. When calculating the projected spatial distributions, we instead use a range spanning the same multiples of the projected stellar half-mass radius of the galaxy, $R_{1/2 M_{\star}}$, which is an average over the three projections of the galaxy along the main axes.

\begin{table}
\centering{
  \caption{Metallicity limits applied to the GC populations as a function of their host galaxy stellar mass: lower metallicity cuts, values used to split between metal-poor and metal-rich objects, and upper metallicity cuts.}
  \label{tab:glxy-upper-fehcuts}
	\begin{tabular}{cccc}\hline\hline
		Galaxy stellar mass & Lower $\feh$ & Bimodality & Upper $\feh$ \\ \hline
		$\log_{10}(M_{\star}/\msun)$ & dex & dex & dex \\ \hline
		8.0 - 8.5 & -2.5 & -1.2 & -1.0 \\
		8.5 - 9.0 & -2.5 & -1.2 & -1.1 \\
		9.0 - 9.5 & -2.5 & -1.2 & -0.8 \\
		9.5 - 10.0 & -2.5 & -1.1 & -0.5 \\
		10.0 - 10.5 & -2.5 & -1.0 & -0.5 \\
		10.5 - 11.0 & -2.5 & -0.9 & -0.5 \\
		11.0 - 11.5 & -2.5 & -0.8 & -0.3 \\
		\hline \hline
	\end{tabular}}
\end{table}

Additionally, we define our GC populations as those clusters that are more massive than $M\geq 10^5~\msun$ at the present day, and that have metallicities above $\feh = -2.5$. As discussed in detail by \citet{kruijssen19a}, the lack of a model describing the cold phase of the interstellar medium in the \eagle galaxy formation model leads to an underdisruption of those clusters that orbit the longest within the gas-rich disk of their host galaxy. In order to prevent the inclusion of artificially underdisrupted clusters, we apply an upper metallicity threshold to our GC populations. This metallicity cut depends on the stellar mass of the host galaxy as more massive galaxies enrich faster, and the cut corresponds to the metallicity at which the median age-metallicity relation of that mass bin starts to saturate \citep[see fig.~1 in][]{horta21}. We expect that clusters with higher metallicities spend more time orbiting in their natal environments, and so they are more affected by underdisruption. We list the upper metallicity cuts applied at each galaxy stellar mass bin in the right side column of Table~\ref{tab:glxy-upper-fehcuts}, and we consider this definition as our fiducial metallicity cut. We compare the mass-dependent upper metallicity cuts to the observed peaks of the metallicity distribution of metal-poor and metal-rich GCs in Virgo \citep[fig.~14 in][]{peng06}, and we find that our upper metallicity limits overlap with the metal-rich peaks. This suggests that our metallicity cut leads to fiducial GC populations that lack about half of their metal-rich objects. We apply this set of criteria to the \citet{harris96} catalogue of GCs \citep[second version,][]{harris10}, and we find that the Milky Way contains 51 objects that correspond to our GC definition.

\begin{figure}
\centering
\includegraphics[width=\hsize,keepaspectratio]{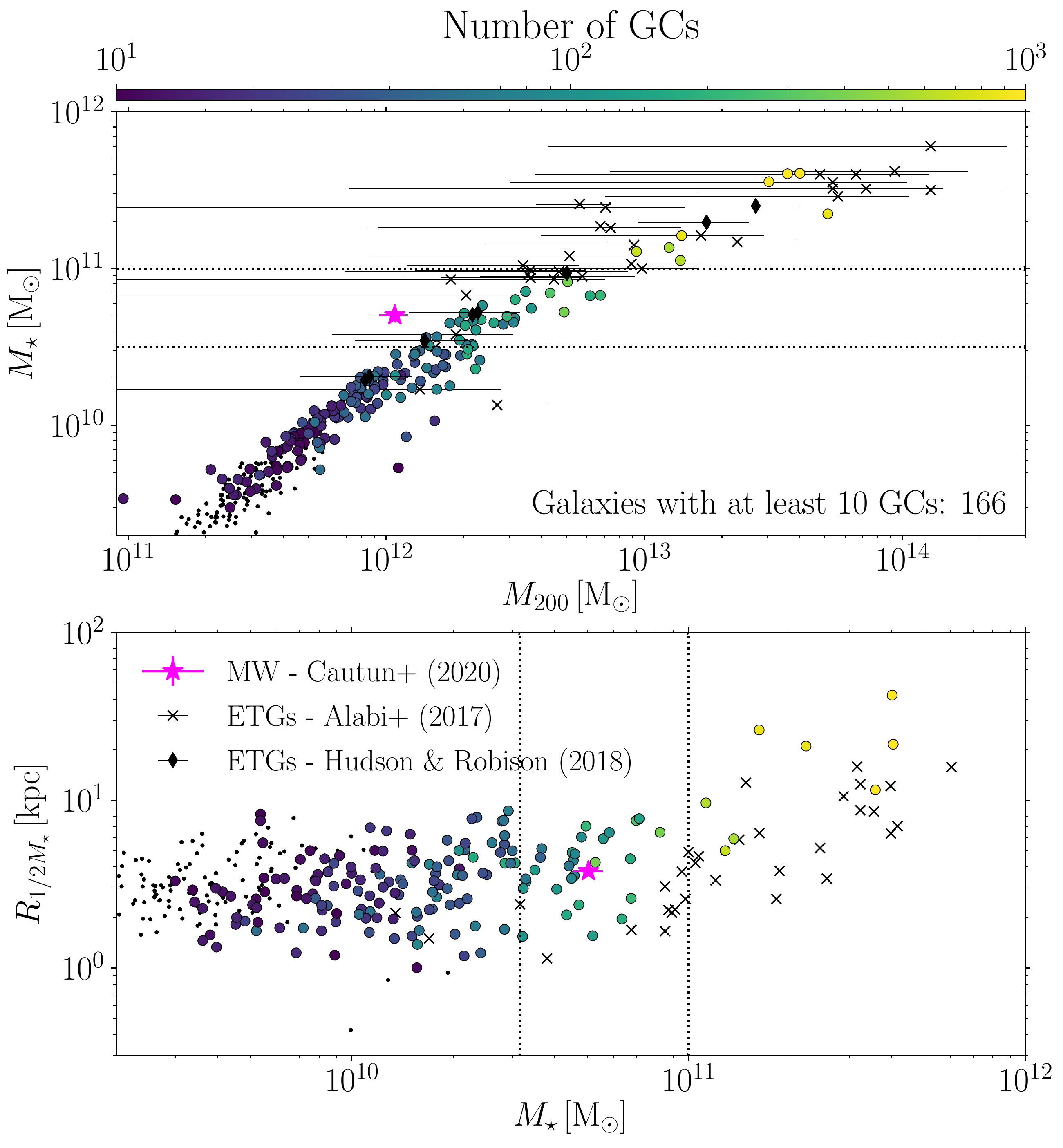}
\caption{\label{fig:stellar-halo-mass} \textit{Top:} Stellar masses of the central galaxies as a function of halo mass. \textit{Bottom:} Projected stellar half-mass radii as a function of the stellar mass of the central galaxies. Small circles show central galaxies with at least 10 GCs within a galactocentric radius in the range $[1,15]\times R_{1/2M_{\star}}$ in the fiducial metallicity cut, and they are colour-coded by the number of GCs they host. Simulated galaxies with smaller GC populations are indicated as black points. The magenta star with errorbars corresponds to the Milky Way \citep{cautun20}. The black crosses with errorbars show the sample of ETGs from \citet{alabi16,alabi17}, and the black diamonds with errorbars correspond to the sample of ETGs from \citet{hudson18}. The thin dotted black lines show the galaxy stellar mass bins used throughout the analysis.}
\end{figure}

Lastly, we restrict our host galaxy sample to contain at least $10$ GCs within the fiducial metallicity and projected ($x$--$y$) radius cuts. The resulting sample consists of $166$ galaxies, and their main characteristics are shown in Fig.~\ref{fig:stellar-halo-mass}. The requirement of the minimum number of GCs elevates the lowest stellar galaxy mass in our sample to $M_{\star}\simeq2.5\times10^9~\msun$\footnote{Irrespective of the number of GCs hosted, there are 258 central galaxies with stellar masses above $2.5\times10^9~\msun$ in the \emosaics volume.}. We summarize in  Table~\ref{tab:glxy-samples} the different galaxy samples used in this work.

\begin{figure*}
\centering
\includegraphics[width=\hsize,keepaspectratio]{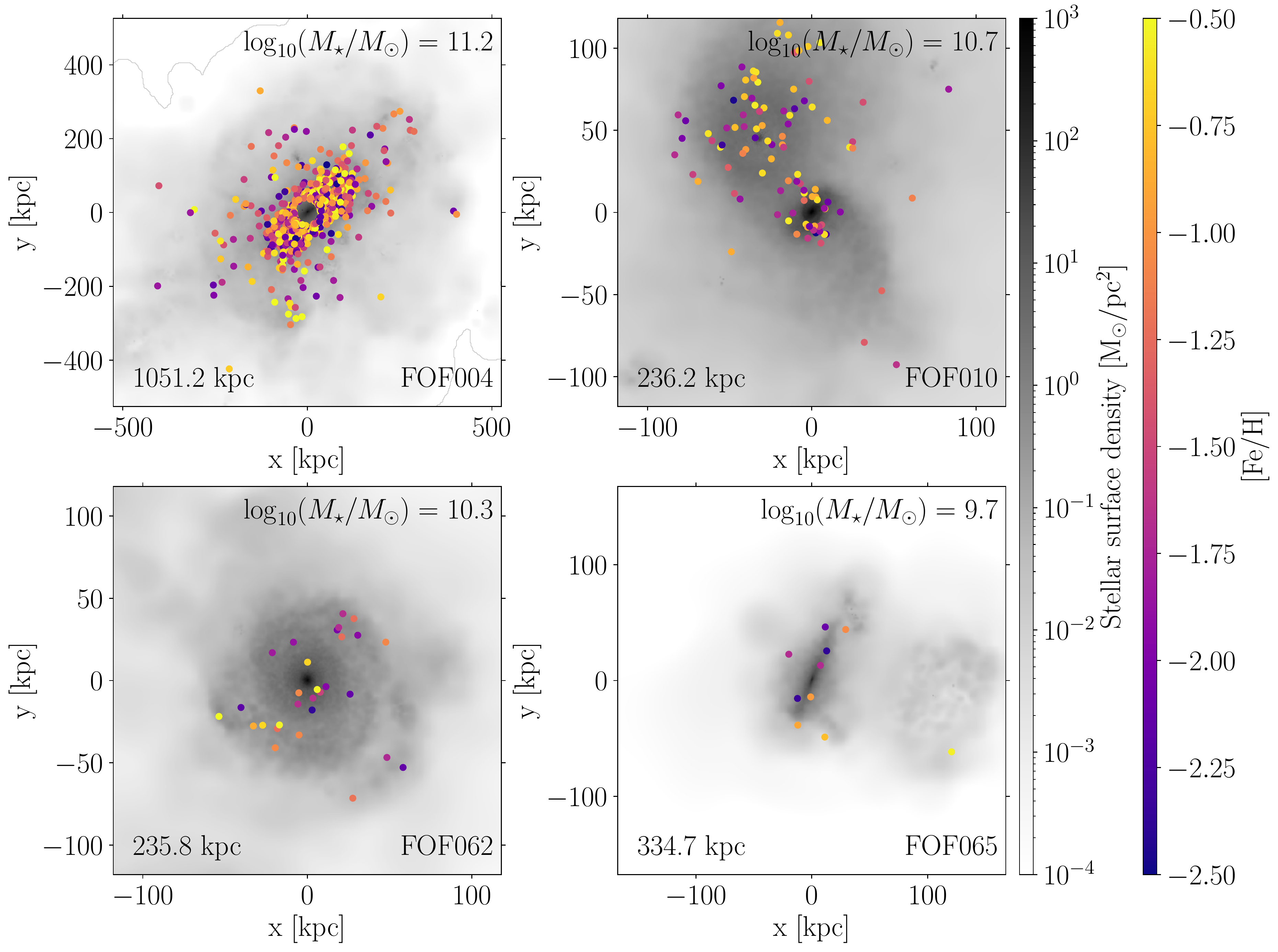}
\caption{\label{fig:spatdistr} Projected stellar surface densities in the $x$--$y$ plane of four galaxies from the \emosaics volume with their GC systems overplotted as coloured data points. These galaxies are a representative sample of the $166$ central galaxies with stellar masses $\log_{10}(M_{\star}/M_{\odot})\geq9.4$, that contain at least $10$ GCs within the fiducial metallicity cut. The width of the panels, indicated in the bottom left corner, is set to $15$ times the stellar half-mass radius of the galaxy, $r_{\rm 1/2 M_{\star}}$, to reproduce the radial range used to select the data (see Sect.~\ref{subsec:select-data}). We indicate the galaxy stellar mass in the top right corner, and the FoF identification number in the bottom right side of each panel.} 
\end{figure*}

We also briefly compare our simulated galaxies to observed objects of which the spatial distribution of their GC system has been studied. We include the Milky Way in this figure, for which we assume a total stellar mass of $M_{\star} = 5.04^{+0.43}_{-0.52}\times 10^{10}~\msun$, a halo mass of $M_{200} = 1.08^{+0.20}_{-0.14}\times 10^{12}~\msun$ and a projected stellar half-mass radius of $ R_{1/2 M_{\star}} = 3.78~\kpc$ \citep{cautun20}. In the high-mass regime, we also include two samples of early-type galaxies (ETGs). The first sample is from the SLUGGS survey presented in \citet{alabi16,alabi17}. For these, we show their effective half-light radius in the bottom panel of Fig.~\ref{fig:stellar-halo-mass}, which tends to be a good tracer of the stellar half-mass radius. The second sample corresponds to the galaxies described by \citet{hudson18}, and are only included in the top panel. Their halo masses are estimated from the stellar to halo mass relation calibrated using weak gravitational lensing \citep{hudson15}, and the errorbars correspond to the intrinsic scatter in the relation,  $0.2~$dex. 

As discussed by \citet{schaye15}, the \eagle galaxy formation model is known to underpredict the stellar masses of galaxies hosted by haloes with $M_{200} \approx 10^{12}~\msun$, so our simulated $L_{\star}$ galaxies have slightly overmassive haloes compared to observed galaxies. Despite this, the most massive simulated galaxies match the stellar-to-halo mass relation of the observed ETG samples. Regarding their sizes, we find that our simulated galaxies are slightly more extended than these observed systems. This can is due to the different morphological types between the observed sample and our simulations (i.e.~ETGs vs any morphological type), as ETGs are observed to be more compact than late-type galaxies \citep[e.g.][]{vanderwel14}. Additionally, the inclusion of intracluster stars in the measurement of the size of the simulated galaxies, which would be excluded by observers, leads to somewhat inflated radii. At the lower mass end, the polytropic equation of state used in the \eagle simulations has been found to produce more extended galaxies relative to observed systems \citep[see][for a detailed comparison across redshift and galaxy mass]{furlong17}, which might also be reflected in the extension of their GC systems. 

We show the diversity of spatial distributions of stars and GCs around a few  selected galaxies in Fig.~\ref{fig:spatdistr}. These central galaxies and their GC populations are a representative subsample of our galaxy selection. It is well-established in observational data, as well as in simulations, that there is a strong trend between the mass of the GC system and the DM halo mass of their host galaxy \citep[e.g.][]{blakeslee97,peng08,georgiev10,harris15,kruijssen15b,harris17c,choksi18,elbadry19,bastian20}. Already from this small subsample, we find the same trend as the observations, i.e.~more massive galaxies host more populous GC systems \citep[see][for a detailed study]{bastian20}. 

The spatial distributions of the GC systems shown in Fig.~\ref{fig:spatdistr} show some intriguing features. The GC systems in some of our galaxies trace stellar debris from recent accretion events (e.g.~FOF$010$), whereas in other galaxies, their GCs preferentially trace their inner galactic structure (e.g.~FOF$065$). If we examine by eye the distributions of subpopulations based on metallicity, we find that metal-poor GCs seem to reside predominantly in the outer regions in some galaxies (e.g.~FOF$062$), whereas both subpopulations are well mixed at all radii in other galaxies (e.g.~FOF$004$). In the next section, we explore in more detail the radial profiles of stars and GCs.

\begin{table*}
\centering{
  \caption{Summary of the samples of simulated central galaxies from the \emosaics volume used in this work. From left to right, columns indicate: the name of the sample, the selected projection, the metallicity range applied to the stellar and GC populations, the total number of galaxies, the number of galaxies per galaxy stellar mass bin, the smallest size of its GC systems, the median size of the GC populations, and the median size per galaxy stellar mass bin. All samples have galaxy stellar masses $M_{\star} \geq 2.5\times10^{9}~\msun$, and the edges of the galaxy stellar mass bins are $\log_{10}(M_{\star}/M_{\odot}) = [9.4, 10.5, 11, 12]$ (as indicated in Fig.~\ref{fig:stellar-halo-mass}). The metallicity cuts are applied as a function of the stellar mass as described in Table~\ref{tab:glxy-upper-fehcuts}. The sample without a metallicity cut (last row of the first block) applies only to stellar populations, and not to GCs.}
  \label{tab:glxy-samples}
	\begin{tabular}{lccccccccccc}\hline\hline
		Sample & Projection & $\feh$ range [dex] & Total $N_{\rm glxs}$ & \multicolumn{3}{c}{$N_{\rm glxs}$ per bin} & $N_{\rm GCs, min}$ & Median $N_{\rm GCs}$ & \multicolumn{3}{c}{Median $N_{\rm GCs}$ per bin} \\ \hline
		Fiducial & $x$--$y$ & Lower -- Upper & 166 & 130 & 28 & 8 & 10 & 31 & 21 & 118.5 & 848.0\\
		Metal poor & $x$--$y$ & Lower -- Bimodality & 166 & 130 & 28 & 8  & 1 & 18 & 12 & 63.5 & 584.5 \\
		Metal rich & $x$--$y$ & Bimodality -- Upper  & 166 & 130 & 28 & 8  & 1 & 12 & 10 & 50.5 & 315.5 \\
		No metallicity cut & $x$--$y$ & -- & 166 & 130 & 28 & 8  & -- & -- & -- & -- & --  \\ \hline
		Fiducial 3D & 3D & Lower -- Upper  & 164 & 128 & 28 & 8 & 8 & 29 & 22 & 120 & 899 \\
		All & 3D & Lower -- Upper & 258 & 222 & 28 & 8 & 0 & 1 & 12 & 120 & 899 \\ 
		\hline \hline
	\end{tabular}}
\end{table*}

\section{Radial distributions of stars and GCs}\label{sec:spatdistr}

In this section, we explore the radial distributions of stars and GCs around central galaxies from the \emosaics simulations.

\subsection{Calculating the radial profiles}\label{subsec:calc-rad-prof}

In order to characterize the spatial distributions of GCs, stars, and DM around the selected central galaxies from the \emosaics volume, we calculate their azimuthally-averaged radial profiles. We determine the GC, stellar and DM profiles in the $258$ central galaxies more massive than $M_{\star}\geq2.5\times10^9~\msun$ (this corresponds to the `All' galaxy sample, see Table~\ref{tab:glxy-samples}). We estimate the stellar and GC radial profiles in three dimensions, as well as the projected profiles for different galaxy orientations. By contrast, since we are interested in inferring the DM profile from the GC populations, we only measure the spherical density profile of the DM halo.

First, we determine the number density profile of GCs, $n(r)$, in $10$ logarithmically-spaced shells with $r\in[1,15]\times r_{\rm 1/2M_{\star}}$ in each central galaxy, as well as the spherical density profile of stars, $\rho(r)$, within the same radial range. We then project each of our central galaxies along three different orientations: face-on, in the $x$--$y$ plane, and edge-on\footnote{We rotate the galaxies such that the angular momentum vector of the stars bound to the galaxy becomes parallel or perpendicular to the $z$-axis for the face-on and edge-on projections, respectively. The $x$--$y$ projection is based on the coordinates of the volume, and it effectively leads to random orientations of the galaxies in our sample.}. For each of these orientations, we calculate the surface number density profile of GCs, $n(R)$, in $10$ bins evenly-spaced in logarithmic radius $R\in[1,15]\times R_{\rm 1/2M_{\star}}$, and the surface density profiles of the stars, $\Sigma(R)$, within the same radial range. We determine these profiles for the fiducial metallicity cut, and also for three other metallicity subpopulations that we define based on the galaxy mass-dependent metallicity cuts in Table~\ref{tab:glxy-upper-fehcuts}. We use the middle point between the peaks of metal-poor and metal-rich GCs in Virgo \citep{peng06} to determine the metallicity that separates both subsamples. This selection describes the metal-poor subpopulations (i.e.~objects with $\feh\geq-2.5$ and less metal-rich than the mass-dependent bimodality cut from Table~\ref{tab:glxy-upper-fehcuts}), the metal-rich subpopulations (i.e.~objects with metallicities within the mass-dependent bimodality and upper metallicity cuts from Table~\ref{tab:glxy-upper-fehcuts}), and the entire population (i.e.~without any metallicity restriction). We summarize the main characteristics of these subsamples in Table~\ref{tab:glxy-samples}.

\begin{figure}
\centering
\includegraphics[width=\hsize,keepaspectratio]{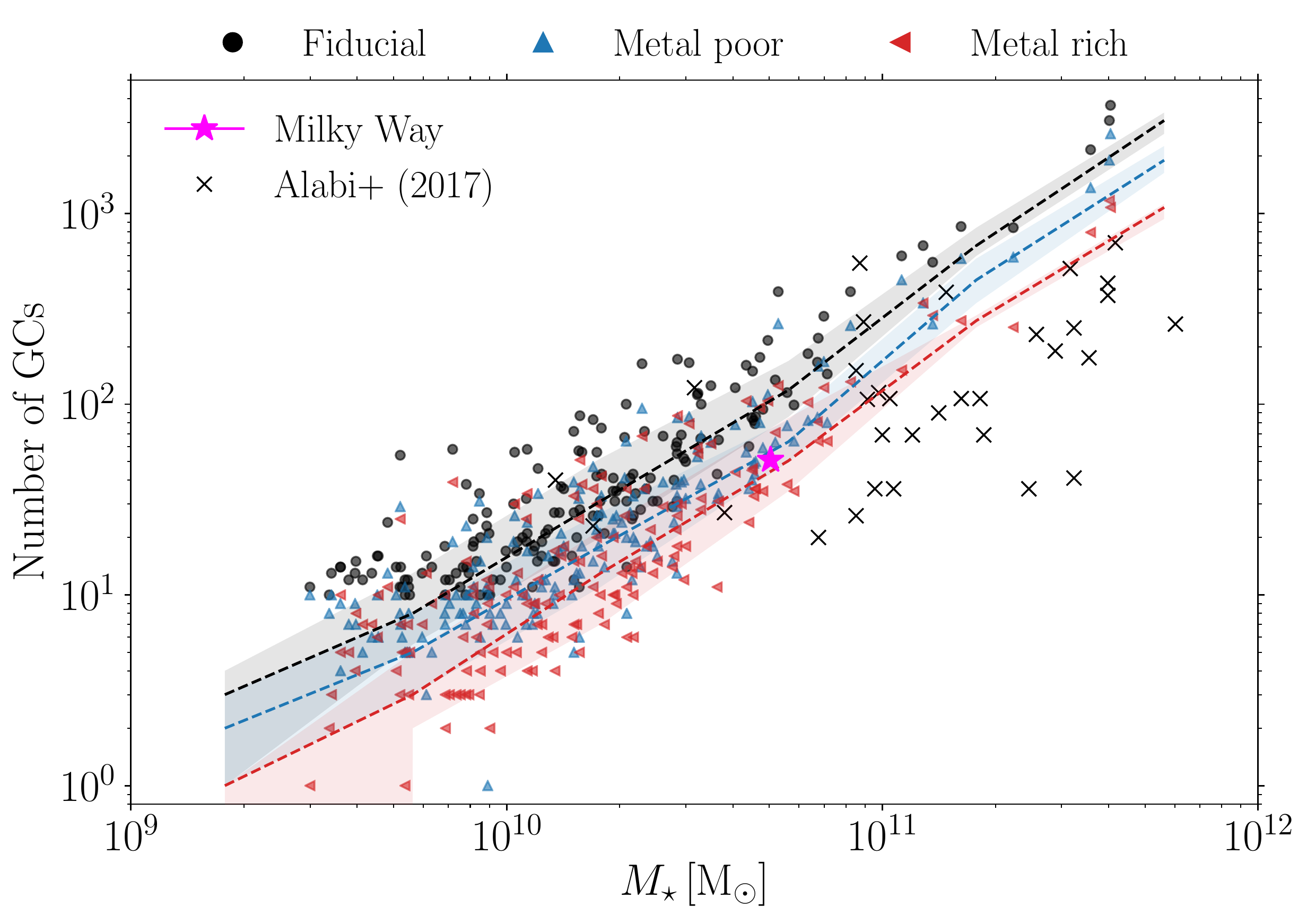}
\caption{\label{fig:num-gcs-feh} Number of GCs in each metallicity subpopulation of GCs within the radial range considered, $R\in[1,15]\times R_{1/2 M_{\star}}$, as a function of galaxy stellar mass in the 166 central galaxies from the \emosaics volume. Metallicity subpopulations are indicated by different small coloured markers, as stated in the legend. The dashed lines with shaded regions show the median and $25$--$75$th percentiles of the overall galaxy sample with at least 1 GC within the radial range. The magenta star corresponds to the Milky Way \citep{harris10,cautun20}, and the black crosses show the number of GCs with kinematic data in the sample of ETGs from \citet{alabi17}. The requirement for galaxies having at least 10 GCs within the fiducial metallicity implies that metallicity subpopulations can be as small as one object (see Table~\ref{tab:glxy-samples}).} 
\end{figure}

We show in Fig.~\ref{fig:num-gcs-feh} the resulting number of GCs in each metallicity subpopulation when projecting the galaxies onto the $x$--$y$ plane. The requirement for galaxies to have at least 10 GCs in the fiducial metallicity cut implies that, at low galaxy masses, the metallicity subpopulations can be as small as a single object (see Table~\ref{tab:glxy-samples}). We find that the number of GCs steeply increases towards more massive galaxies, as expected from the increasing mass of the GC system towards more massive haloes \citep[see e.g.][]{peng08,georgiev10,harris15,harris17c,bastian20}. Our simulated GC populations are dominated by the metal-poor objects across our galaxy mass range, whereas observations find that the fraction of metal-rich GCs increases with galaxy mass \citep[e.g.][]{peng06,harris15}. As discussed in Sect.~\ref{sec:emosaics}, the mass-dependent upper metallicity cut introduced to mitigate contamination from underdisruption overlaps with the peak of metal-rich GCs in Virgo \citep{peng06}. This implies that our metal-rich GC subpopulations miss about half of their objects, and prohibits us from doing accurate comparisons of the relative contribution of each metallicity subpopulation.

Lastly, we calculate the spherical density profile of the DM halo, $\rho_{\rm DM}(r)$, by binning its mass distribution in 32 shells evenly-spaced in logarithmic radius between $[0.05, 1]$ times the virial radius of the halo, $r_{\rm 200}$ \citep[see][for a discussion of the radial range]{neto07}. We discuss the stellar and GC profiles in Sect.~\ref{subsec:radprof-emosaics}, and we further characterize all the radial profiles by fitting analytical distributions in Sect.~\ref{sec:fit-profiles}. 

\subsection{Radial profiles of stars and GCs in E-MOSAICS}\label{subsec:radprof-emosaics}

\begin{figure*}
\centering
\includegraphics[width=\hsize,keepaspectratio]{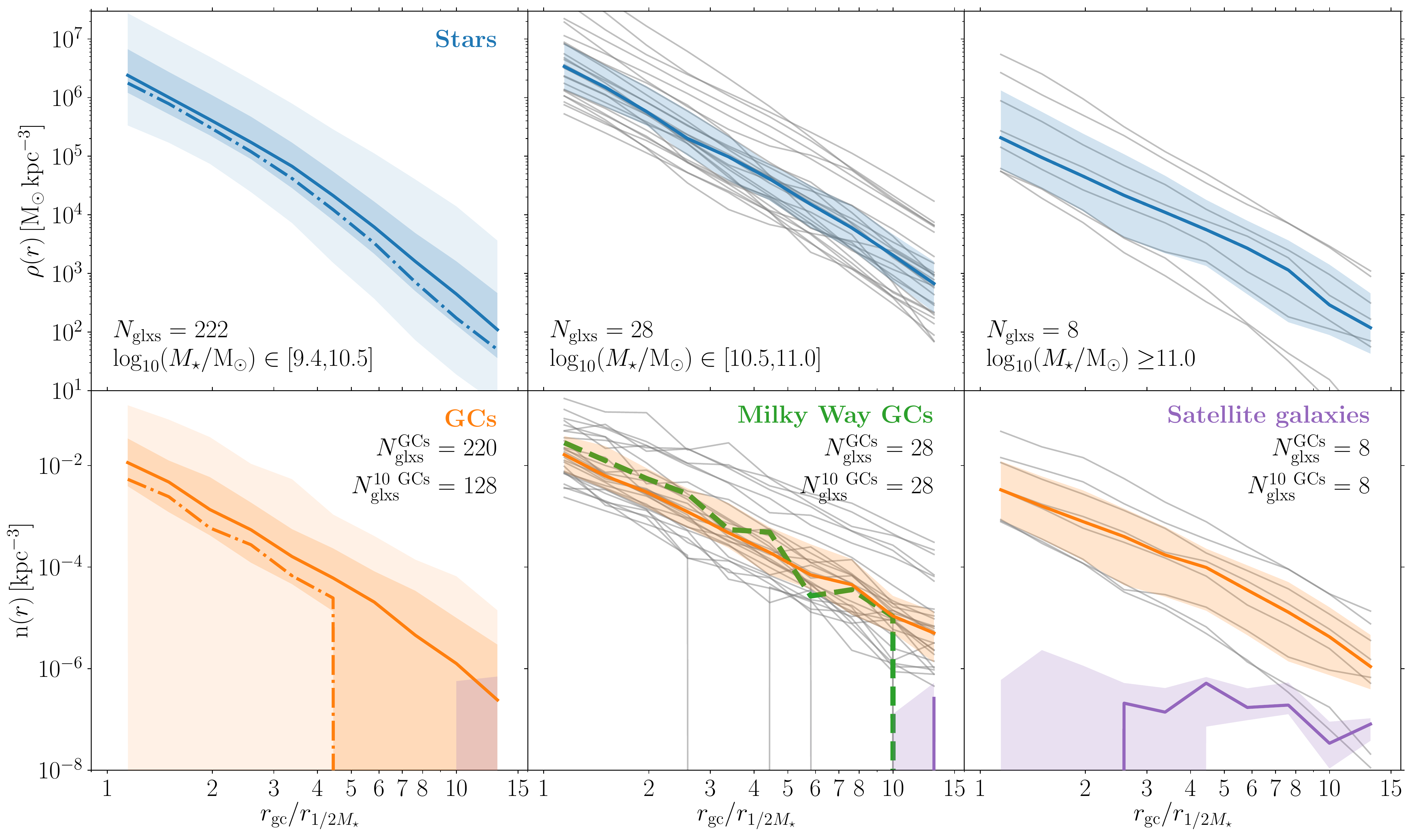}
\caption{\label{fig:r3d} Spherical radial profiles of stellar density (\textit{top row}) and GC number density (\textit{bottom row}) in units of the 3D stellar half-mass radius of the galaxy. Each column corresponds to a different galaxy stellar mass bin for the central galaxies from the \emosaics volume. The number of galaxies within the mass bin is indicated in the bottom left corner of each panel in the top row ($N_{\rm glxs}$), and the number of galaxies hosting GCs within the mass bin ($N_{\rm glxs}^{\rm GCs}$), and those with at least 10 GCs ($N_{\rm glxs}^{\rm 10~GCs}$), are indicated in the top right corner of each panel in the bottom row. Blue and orange solid lines and shaded regions indicate the median and $25$--$75$th percentiles of stars and GCs, respectively, for galaxies with at least 10 GCs (`Fiducial 3D' sample in Table~\ref{tab:glxy-samples}). Dash-dotted blue and orange lines indicate the median for all galaxies (`All' sample in Table~\ref{tab:glxy-samples}). Thin grey lines in the middle and right columns correspond to the stellar and GC radial profiles in individual galaxies within the corresponding mass bins, whereas the lighter shaded regions in the left column indicate the $5$--$95$th percentiles. The green dashed line in the middle bottom panel corresponds to the radial distribution of Galactic GCs that match the same criteria as applied to the simulated GCs. The purple solid lines and shaded regions in the bottom row correspond to the median and $25$--$75$th percentiles of the number density profile of satellites around the sample of central galaxies with at least one satellite. See the text for more details on each sample. Populations of GCs are more numerous and more spatially extended than satellite galaxies, which makes them good tracers of the galactic outskirts.}
\end{figure*}

We show in Fig.~\ref{fig:r3d} the spherical radial profiles of stars and GCs around the sample of 258 central galaxies more massive than $M_{\star}\geq2.5\times 10^{9}~\msun$. The median stellar density profile (top row in Fig.~\ref{fig:r3d}) changes from a broken power-law in the lowest galaxy mass bin to a single power-law at higher masses. This indicates that less material is deposited in the outer parts of low-mass galaxies during their assembly relative to higher mass galaxies. A similar conclusion is reached by \citet{font11} when examining the growth of stellar haloes in a large sample of $L_{\star}$ galaxies from the GIMIC simulations. The authors find that the transition between the halo being dominated by in-situ stars in the inner region to being mostly accreted in the outskirts produces a similar change in the slope of the stellar surface densities as seen in our simulated galaxies. We also find that the median profiles of galaxies with at least 10 GCs are higher than the median profile over all galaxies, suggesting that lower stellar surface brightness galaxies have been less able to form populous GC systems. This difference increases at large distances, such that fainter stellar haloes host smaller GC populations.

Across the galaxy stellar mass bins, we find that the median radial profiles of stars and GCs become shallower for more massive galaxies. Since we focus our analysis on the populations around central galaxies, this does not correspond to the presence of objects currently linked to satellite galaxies. Instead, this reflects that the growth of more massive galaxies is dominated by the accretion of large numbers of  satellites that can deposit their stellar and GC populations further out. By contrast, lower mass galaxies grow mostly due to in-situ star formation and are predicted to have low fractions of accreted stars \citep[e.g.][]{abadi06,rodriguez-gomez16,qu17,clauwens18,behroozi19,choksi19b,davison20,remus21}, which leads to steeper radial profiles for their stellar and GC populations. This result is consistent with the findings of  \citet{pillepich14,pillepich18} for the stellar haloes in \illustris and \illustrisTNG, but in this work we extend the analysis to the GC populations in the \emosaics project. Given the brighter luminosities of GCs relative to the diffuse stellar component in the outskirts of galaxies, they are more useful tracers of the outer matter distribution of their host galaxy. In this study, we explore if the radial profiles of GCs can trace the structure of the DM halo and the assembly history of their host galaxies.

The radial number density profiles of GCs (bottom row in Fig.~\ref{fig:r3d}) are noisier in lower-mass galaxies, due to the smaller number of GCs hosted by those systems (see Fig.~\ref{fig:num-gcs-feh}). In the lowest galaxy mass bin, $40~$per cent of the simulated galaxies have fewer than 10 GCs in the fiducial cut, and $2$ of them have no GCs at all. These smaller GC systems tend to be more concentrated, with the majority being within ${\sim}5$ times the stellar half-mass radius. Using the catalogue of \citet{harris10}, we estimate the number density profile of GCs in the Milky Way, and this is shown in the middle panel of Fig.~\ref{fig:r3d}. We find that there is good agreement, within the observed galaxy-to-galaxy scatter, between the Galactic GCs and our simulated populations \citep[also see][who reported this for the 25 Milky Way-mass zoom-in simulations of \emosaics]{kruijssen19a}. 

Finally, we compare the spatial distributions of GCs with those of satellite galaxies. We include the median number density profile of satellites around the sample of central galaxies in the bottom row of Fig.~\ref{fig:r3d}. For this, we consider only central galaxies with at least one bound satellite within the radial range considered, i.e.~$[1,15]$ times the stellar half-mass radius. We only consider satellite galaxies more massive than $M_{\star} \geq 2.2\times10^7~\msun$, as this galaxy stellar mass limit ensures that the satellites are resolved by at least $100$ stellar particles. We find that satellite galaxies preferentially reside in the outer regions across the galaxy mass range\footnote{Applying an explicit mass distribution tensor approach to the \eagle simulations, \citet{velliscig15b} find that satellite galaxies lie in anisotropic distributions in which they preferentially reside along the major axis of the central galaxy. In contrast, our azimuthally-averaged radial analysis prevents us from drawing similar conclusions.}. More massive galaxies host a larger number of satellite galaxies, with the most massive haloes ($M_{\star}\geq10^{11}~\msun$) containing about $\sim100$ satellites. This is an order of magnitude smaller than their corresponding GC systems, even in the outermost bin at $r \sim 15r_{1/2M_*}$ (see Fig.~\ref{fig:num-gcs-feh}). In contrast with the satellite population, galaxies host more numerous GC systems, thus being more suitable tracers of the mass distribution in the galactic outskirts.

\begin{figure}
\centering
\includegraphics[width=\hsize,keepaspectratio]{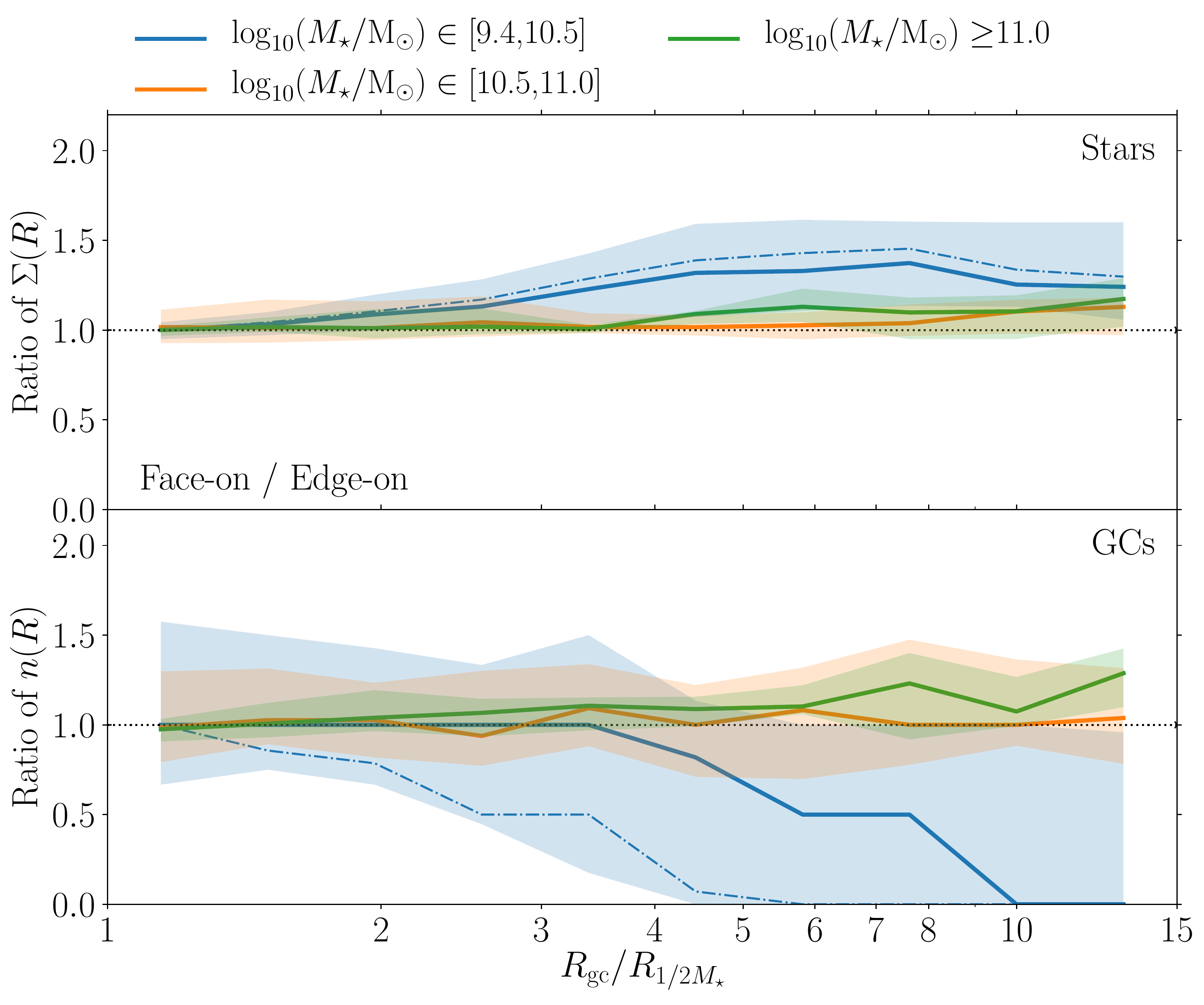}
\caption{\label{fig:r2d-ratios} Ratios of the face-on relative to the edge-on radial profiles for the spatial distributions considered: projected stellar surface density profile (\textit{top panel}) and projected number density profile of GCs (\textit{bottom panel}). Each line corresponds to the median ratio for a given galaxy stellar mass bin. Solid lines and shaded regions indicate the median ratios and their $25$--$75$th percentiles for galaxies with at least 10 GCs, whereas dash-dotted lines show the ratios when considering all the galaxies within the mass bin. The dash-dotted lines can only be distinguished from the solid lines in the lowest mass bin. Populations of GCs in massive galaxies are close to being spherical, whereas lower galaxy mass bins suffer from the stochasticity of low number of objects, specially at large galactocentric radius.}
\end{figure}

Next, we explore the shapes of the spatial distributions of stars and GCs. For this, we project our sample of galaxies along three different orientations (i.e.~face-on, $x$--$y$, and edge-on), and we determine the azimuthally-averaged projected radial profiles of stars and GCs. Then, we calculate the ratio between the radial profiles over the face-on and edge-on projections. If the ratio is close to one, that population has a nearly spherical distribution. We show in Fig.~\ref{fig:r2d-ratios} the median ratios between the face-on and the edge-on projections over three galaxy mass bins. 

The median ratios of the stellar and GC projected profiles are remarkably close to unity, implying that stars and GC systems are, on average, close to being spherical. The stellar surface density ratio shows a small deviation from sphericity at large radii in the lowest galaxy mass bin, with the face-on profile becoming larger than the edge-on profile. This suggests that, at large radii, the stellar haloes of low mass galaxies resemble slightly oblate spheroids, whereas they tend to be spherical in more massive galaxies. The deviation from sphericity is even more pronounced in the GC populations hosted by the lowest galaxy mass bin. The decreasing median ratio towards large radii suggests these populations might be slightly prolate. However, since the number of objects quickly drops at the low galaxy mass end, the distributions are more stochastically sampled. This leads to a larger scatter in the ratios of the radial profiles, particularly at large galactocentric radius. When comparing the trends of the ratios of all GC populations (dash-dotted blue line) relative to those of populations that host at least $10$ objects (solid blue line), we find evidence that the decreasing trend is dominated by sampling noise. From this, we assume that the stellar and GC populations are nearly spherical, and so they can be well approximated by an azimuthally-averaged description in projection. Therefore, the rest of the paper only considers the projected radial profiles obtained over the random $x$--$y$ plane. This allows us to reproduce the random distribution of orientation in observed extragalactic systems, and will simplify the comparison of our projected radial profiles with observational data. Additionally, from here on we restrict our analysis to the sample of 166 central galaxies that contain at least 10 GCs in the fiducial metallicity cut.

Observations of GC populations have found that metal-poor objects tend to be more radially extended than the metal-rich subpopulations in a variety of galactic environments \citep[e.g.][]{zinn85,rhode04,bassino06,caldwell11,faifer11,pota13,kartha14,cho16,kartha16,hudson18}. Differences in the spatial distributions of metal-rich and metal-poor GCs have been suggested to result from their formation in different galactic environments. For example, metal-poor GCs that form in the early Universe in low mass satellites are later accreted onto the outer regions of massive galaxies. In contrast, metal-rich GCs reside in the inner part of the galaxy either because they form in-situ in the massive galaxy, or because they are accreted from massive satellites \citep[e.g.][]{brodie06}. This scenario suggests that the subpopulations of GCs in the outskirts of galaxies can be good tracers of the structure and assembly of their host galaxies. 

\begin{figure*}
\centering
\includegraphics[width=\hsize,keepaspectratio]{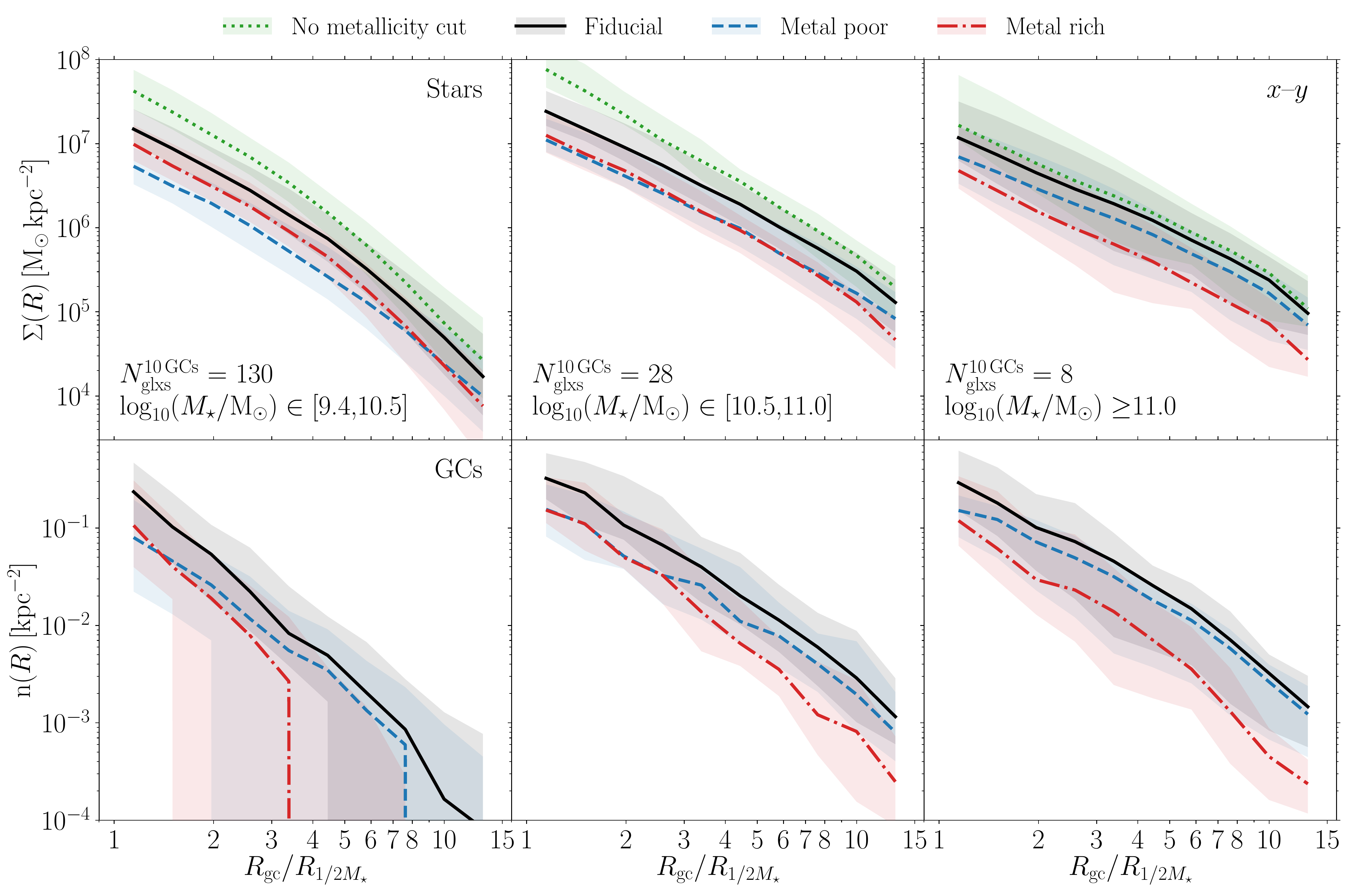}
\caption{\label{fig:r2d-feh} Median projected radial distributions of metallicity subpopulations of stars and GCs around the central galaxies from the \emosaics volume: projected stellar surface density profile (\textit{top row}) and projected number density profile of GCs (\textit{bottom row}). The spatial distributions are obtained assuming random galaxy orientations (using the $x$--$y$ plane of the simulated volume), and the main characteristics of these galaxy samples are described in Table~\ref{tab:glxy-samples}. Each column corresponds to a different galaxy stellar mass bin, and the number of galaxies within the mass bin with at least $10$ GCs ($N_{\rm glxs}^{\rm 10~GCs}$) is indicated in the bottom left side of the top row. Coloured lines with shaded regions correspond to medians and the $25$--$75$th percentiles of different metallicity subpopulations. Only central galaxies with at least 10 GCs within the fiducial metallicity cut are considered. Metal-poor GC subpopulations tend to dominate the galactic peripheries across the galaxy mass range, whereas the inner parts of the galaxies are dominated by the metal-rich objects.}
\end{figure*}

We now investigate whether our simulated stellar and GC populations also show similar differences in their radial distributions when considering different metallicity cuts. For this, we show in Fig.~\ref{fig:r2d-feh} the median projected stellar surface density and number density profiles of stars and GCs, respectively, in the top and bottom rows. We find that metal-poor subpopulations of stars and GCs become the dominant subpopulations at large distances ($\gtrsim5\times R_{1/2 M_{\star}}$) with increasing galaxy mass. In the lowest galaxy mass bin, the broken power-law shape observed in the three-dimensional profiles (Figs.~\ref{fig:r3d}) is driven by the metal-rich subpopulations. In low-mass galaxies, the majority of the metal-rich material in the outskirts is the result of in-situ star formation, which dominates the growth of the stellar halo \citep[e.g.][]{rodriguez-gomez16,behroozi19}, and is therefore more concentrated. More massive galaxies build up their metal-rich stellar haloes both from in-situ star formation \emph{and} from the accretion of massive satellites, which themselves have more metal-rich stars than low mass satellites \citep{ma16}, and so their radial profiles have shallower slopes. This result echoes the findings of \citet{font11}, who find that in-situ metal-rich stellar populations dominate the inner regions of the stellar haloes surrounding $L_{\star}$ galaxies, whereas the outer regions tend to be mainly accreted and metal-poor. We explore the role of the formation mode below.

For any given galaxy mass bin we find that metal-poor objects tend to have shallower radial profiles relative to the metal-rich subpopulations. When considering the overall galaxy sample, the radial profiles of both subpopulations become shallower in more massive galaxies, which is a hint of the assembly history of their galaxies \citep{abadi06,pillepich14,pillepich18}. We further quantify and discuss the projected radial profiles of the metallicity subpopulations in Sect.~\ref{sec:fit-profiles}. The difference in the radial profiles of metal-poor and metal-rich GCs is more prominent in the lowest galaxy mass bin, and a similar trend is seen in the stellar populations. This suggests that the radial profiles of metallicity subpopulations in this galaxy mass bin ($\log_{10}(M_{\star}/M_{\odot})\leq 10.5$) might be more sensitive to the different origin of these objects, which we explore further in Sect.~\ref{subsec:prof-stars-gcs}. Finally, we also find that the metal-rich GCs tend to dominate the inner galaxy out to $4$--$6 \times R_{1/2M_{\star}}$, with metal-poor objects becoming more numerous in the outskirts. This is in good agreement with observed GC populations \citep[e.g.][]{caldwell11,pota13}, and is further discussed in the next section, where we make quantitative comparisons.

\section{Characterizing the radial profiles}\label{sec:fit-profiles}

In this section, we characterize the projected radial profiles of stars and GCs, as well as the spherical profiles of the DM haloes by fitting analytical profiles with different functional forms. We use the populations of objects around the $166$ central galaxies (with $M_{\star}\geq2.5\times10^9~\msun$) that contain at least $10$ GCs within the fiducial metallicity cut (first block of samples in Table~\ref{tab:glxy-samples}). 

\subsection{Stars and GCs}\label{subsec:prof-stars-gcs}
We follow the same procedure to fit the projected azimuthally-averaged radial profiles of stars and GCs. First, we calculate the projected galactocentric radius $R$ of stars and GCs within the range $R\in[1,15]\times R_{1/2 M_{\star}}$ in each of the 166 central galaxies. To calculate the normalizations of the radial profiles, we also determine the total mass in stars and the total number of GCs within the radial range. We then use a maximum likelihood method to find the combination of parameters that maximize the likelihood of the system studied. This method avoids having to bin the data, which can lead to large errors in the estimation of the parameters when a small number of objects is considered. The likelihood of our systems is defined as
\be 
	\ln \mathcal{L}(R) = \sum_i^{N} \ln \mathcal{P}(R_{i}),
\ee
where $N$ is the number of objects considered. The probability $\mathcal{P}$ that an object $i$ is at its projected radius $R_{i}$ given a profile $f(R)$ is 
\be 
	\mathcal{P} (R_{i}) = 2\pi R_{i} f(R_{i}).
\ee
Observational studies suggest that the radial profiles of GC populations can be well characterized with both power-law distributions and S\'ersic-like profiles \citep{sersic63,sersic68}, i.e.~a power-law shape in the outer regions that flattens in the center \citep[e.g.][]{faifer11,kartha14,cho16,hudson18}. The radial range considered in this work does not include the central stellar half-mass radii of the galaxies, so we use the maximum likelihood estimation to characterize the populations of stars and GCs assuming two different functional forms: a power-law function, and a de Vaucouleurs profile (i.e.~equivalent to a S\'ersic profile with a slope $n = 4$). These functions are generally used in the literature to describe the radial profile of GCs \citep[e.g.][]{hudson18}.

We describe the power-law distribution as,
\be
	f(R) = f_{\rm e} R^{-\alpha},
	\label{eq:power-law}
\ee
where $R$ is the projected galactocentric radius of stars and GCs. We aim to find the slope $\alpha$ that maximizes the likelihood of the system. Note that the slope $\alpha$ is constant over the radial range considered. For each subpopulation studied, we calculate the normalization $f_{\rm e}$ of the projected profiles using the following analytical expression:
\be
	f_{\rm e} = \dfrac{2-\alpha}{2\pi (R_{\rm max}^{2-\alpha} - R_{\rm min}^{2-\alpha})},
\ee
where the radii $R_{\rm min}$ and $R_{\rm max}$ correspond to the inner and outer edges of the radial range considered. In the case that the subpopulations of GCs do not cover the entire radial range, we modify these radii to be the smallest and largest radii of GCs in that subpopulation, respectively. 

In order to obtain an estimate of the effective size of GC populations, we also fit de Vaucouleurs profiles to the subpopulations of stars and GCs, 
\be
	f(R) = f_{\rm e} \exp\left\{-b_{4}\left[{\left(\dfrac{R}{R_{\rm e}}\right)}^{1/4}-1\right]\right\},
\ee
where $b_4 = 7.669$ \citep{graham05}, $R_{\rm e}$ is the effective radius, and $f_{\rm e}$ is the density at that radius. Relative to using a standard S\'ersic profile, in which both the slope and the effective radius are free parameters, we find that fixing the slope is crucial to avoid noisy fits because it reduces the degrees of freedom. For a given effective radius, $R_{\rm e}$, the normalization of this profile can be calculated as,
\be
\begin{split}
	f_{\rm e} =& \dfrac{b_{4}^8 e^{-b_4}}{8\pi R_{\rm e}^2 } \times \\
	&\left\{\gamma\left[8, b_{4} \left(\frac{R_{\rm max}}{R_{\rm e}}\right)^{\frac{1}{4}}\right] - \gamma\left[8, b_{4} \left(\frac{R_{\rm min}}{R_{\rm e}}\right)^{\frac{1}{4}}\right]\right\}^{-1}.
\end{split}
\ee
Both of these functional forms have only one degree of freedom ($\alpha$ and $R_{\rm e}$, respectively), which reduces the noise introduced from overfitting parameters.

Finally, the normalizations $f_{\rm e}$ are multiplied by either the total mass in stars or the total number of GCs within the radial range,
\be
	n(R) = n_{\rm e}R^{-\alpha} \rightarrow n_{e} = f_{\rm e}\times N_{\rm GCs}(R_{\rm min}<R<R_{\rm max}),
\ee
depending on whether the radial profile represents a mass or number density profile, respectively. We repeat this fitting procedure to determine the slopes of the projected radial profiles of the metallicity subpopulations of stars and GCs described in Sect.~\ref{subsec:calc-rad-prof} (using the metallicity limits from Table~\ref{tab:glxy-upper-fehcuts}): fiducial, metal-poor, metal-rich and, only in the case of stars, without a metallicity selection (first block of galaxy samples in Table~\ref{tab:glxy-samples}). To avoid spurious measurements, we only fit profiles to subpopulations that have more than 3 objects.

\begin{figure*}
\centering
\includegraphics[width=\hsize,keepaspectratio]{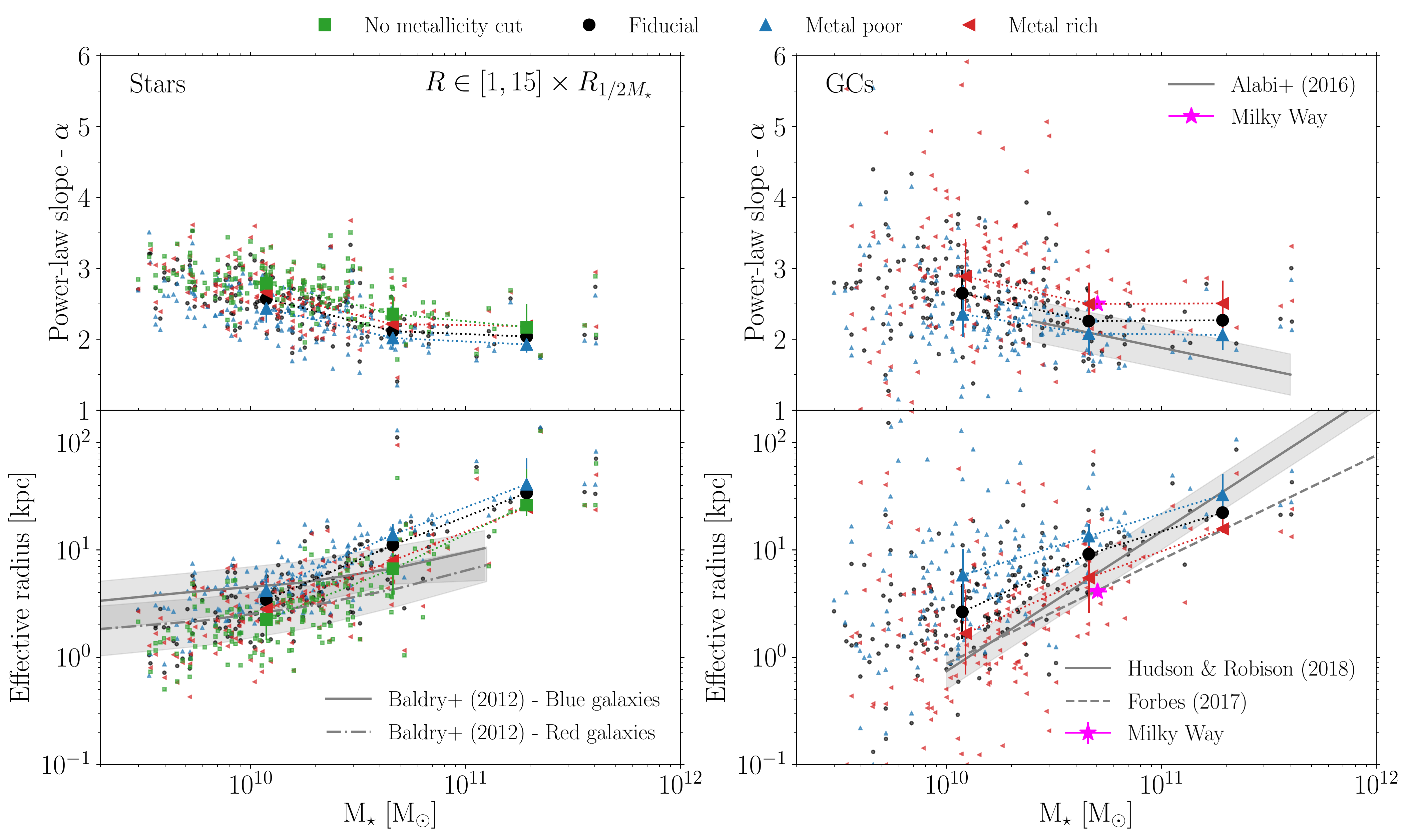}
\caption{\label{fig:fits-stars-gcs} Characterization of the projected radial profiles of different metallicity subpopulations of stars and GCs around central galaxies from the \emosaics volume: fitted power-law slope (\textit{top row}), and effective radius of the de Vaucouleurs profile of each subpopulation (\textit{bottom row}), as a function of galaxy stellar mass. Data points correspond to the 166 central galaxies with at least 10 GCs within the fiducial metallicity cut. Metallicity subpopulations are indicated by different small coloured markers. Coloured markers with errorbars connected by dotted lines show median values and the $25$--$75$th percentiles. The grey solid and dash-dotted lines in the bottom left panel show the median half-light radius of blue and red galaxies from the GAMA survey \citep{baldry12}, respectively, and the grey shaded regions indicate the $16$--$84$th percentiles. The magenta stars with errorbars in the right-hand column correspond to the Milky Way \citep{harris76,wolf10,hudson18,cautun20}. The grey line and shaded region in the top-right panel corresponds to the observational fit obtained by \citet{alabi16}, whereas the grey solid and dashed lines in the bottom-right panel correspond to the observational relations obtained by \citet{hudson18} and \citet{forbes17}, respectively. Massive galaxies host shallower and more radially extended distributions of stars and GCs, and metal-poor subpopulations tend to have shallower and more extended profiles than their metal-rich counterparts.}
\end{figure*}

We maximize the likelihood of each subpopulation to obtain the best-fitting parameters of a given functional form. For that, we use initial guesses for the value of the parameters: the power-law slope is initially set to $\alpha = 1$, and the de Vaucouleurs effective radius is firstly assumed to be the median radius of the subpopulation. In the case of the de Vaucouleurs profile, we bound the radius to be within the range $0.1~\kpc$ to $150\times R_{1/2 M_{\star}}$ to recover sensible parameters. We try different priors for the parameters, and we find that the recovered best-fitting parameters are insensitive to the choice of the initial guesses.

To estimate the quality of the fits, we calculate the likelihood of the best-fitting power-law and de Vaucouleurs profiles of the stellar and GC subpopulations, which we show in Appendix~\ref{app:fits-quality}. Lower mass galaxies with fewer particles have poorer stellar and GC fits. Because of our requirement that galaxies host at least 10 GCs within the fiducial metallicity cut, the metallicity subpopulations in low-mass galaxies can have very low numbers of GCs (see Fig.~\ref{fig:num-gcs-feh}). This implies that the radial profiles of these smaller systems are poorly constrained, and they introduce some scatter when examining the recovered parameters. However, we decide to keep the requirement on the number of GCs in the fiducial cut so that our analysis is based on the same sample of 166 galaxies regardless of the metallicity cut. From the quality analysis, we find that the stellar profiles are always better described by a de Vaucouleurs profile, but the GC populations in lower mass galaxies (up to $M_{\star}\simeq5\times10^{10}~\msun$ for the metal-poor GCs) are better characterized by power-law distributions. 

We show the recovered power-law slopes and de Vaucouleurs effective radii of the stellar and GC subpopulations in Fig.~\ref{fig:fits-stars-gcs}. Focusing on the stellar populations first (left-hand column), we find that the power-law slopes describing their radial profiles are within the range $\alpha=1$--$3.5$, and the values show little scatter. The mild decreasing trend of the slope towards more massive galaxies is the result of the higher accreted fractions in massive galaxies \citep{pillepich14, pillepich18}. The stellar populations have effective radii between $R_{\rm e} = 1$--$50~\kpc$, and they steeply increase towards higher mass galaxies \citep[e.g.][]{shen03,baldry12,lange15}. We include the half-light radius of the red and blue galaxies from the GAMA survey \citep[][bottom-left panel of Fig.~\ref{fig:fits-stars-gcs}]{baldry12}, and we find that the measured stellar effective radii of our simulated galaxies agrees well with these observations.

If we now look at the GC populations (right-hand column in Fig.~\ref{fig:fits-stars-gcs}), we find that these reproduce the same median trends as obtained for the stellar populations. Given that stellar haloes trace the slope of the DM halo \citep{pillepich14}, this similarity suggests that GC populations might also trace the structure of the DM halo of their host galaxy. The large scatter shown by the recovered parameters describing the GC subpopulations is driven by the low number statistics in low mass galaxies ($M_{\star}\leq4\times10^{10}~\msun$). At those stellar masses, our galaxies have a median of $\sim10$--$80$ GCs within the fiducial metallicity cut (see Fig.~\ref{fig:num-gcs-feh}), and the metallicity subsamples can include as few as $3$ objects. 

The power-law slopes describing the surface number density profiles of the GCs populations also show a decreasing trend towards higher mass galaxies. This trend is found across the different metallicity subpopulations considered, and it reproduces extragalactic observations of shallower slopes for GC systems hosted in brighter galaxies \citep[e.g.][]{harris86,kisslerpatig97,ashman98,dirsch05,bekki06b,alabi16,alabi17,hudson18}. We suggest that this is the result of higher mass galaxies hosting larger fractions of accreted GCs \citep[e.g.][]{harris17a}, and we further explore this scenario below.

The slopes of our projected fiducial GC subsamples are in the range $\alpha=1$--$4.5$, and seem to flatten in the very high-mass end. \citet{alabi16} perform a literature compilation of extragalactic systems, and provide a fit to the de-projected slope of the GC spatial profile as a function of the galaxy stellar mass. We project those slopes by substracting one dex to them, $\alpha_{\rm 2D} = \alpha_{\rm 3D} - 1$, and include the relation in Fig.~\ref{fig:fits-stars-gcs} (top-right panel). The shallower simulated slopes relative to the observed galaxies are likely due to our wide radial range, which ignores the inner stellar half-mass radius of the galaxy and extends up to $15$ times that radius. In order to test this idea, we repeat our fitting procedure for different radial ranges. We show in Appendix~\ref{app:diff-radial-range} that the recovered slope over narrower radial ranges shows a better agreement with the observed trend.

When fitting a de Vaucouleurs profile, we also find that the effective radii of the GC subpopulations increase steeply towards more massive galaxies, with the median effective radius of the fiducial GC subpopulation in the range $R_{\rm e}=5$--$30~\kpc$. This increasing trend has been observed in a variety of galactic environments \citep[e.g. see fig.~18 of][]{kartha14}, suggesting that more massive galaxies host more radially extended populations of both stars and GCs. Since the larger extent of the stellar populations in more massive galaxies is mostly due to their accreted origin \citep{abadi06,pillepich14,pillepich18,font20,remus21}, we later explore if the larger size of the GC populations can also be related to a higher fraction of them having formed in accreted satellites. 

In order to compare the increasing effective size of the fiducial GC subpopulations to observed extragalactic systems, we include the observed relations from \citet{hudson18} and \citet{forbes17} in Fig.~\ref{fig:fits-stars-gcs} (bottom-right panel). Both studies use samples of ETGs to study the correlation between the spatial extent of the GC systems and their DM haloes, and they obtain slightly different relations between the size of the GC systems and the effective radii of their galaxies. They both find that, as ETGs grow in mass, their GCs populations grow proportionally in size, in good agreement with what we also see in the simulated populations. \citet{forbes17} complements the sample of ETGs with the GC systems in three ultra-diffuse galaxies (UDGs), and they find that the relation becomes shallower for GC populations in UDGs at galaxy masses $M_{\star}\leq4\times10^{10}~\msun$, which roughly corresponds to the same mass as the change in slope of the galaxy mass--size relation \citep[e.g.][]{shen03,baldry12,lange15}. We observe hints of a similar flattening in the size of the fiducial GC populations of our simulated galaxies in the low-mass regime ($M_{\star}\lesssim2\times10^{10}~\msun$), even though this is the galaxy mass range in which there are low number statistics.

If we now focus on the metallicity subpopulations of GCs for a given galaxy mass, we find that more metal-poor subsamples have shallower radial profiles that are more extended. The median power-law slopes of the metal-poor GC subsamples are in the range $\alpha=2$--$2.3$, whereas the metal-rich objects have median slopes $\alpha\sim2.5$--$2.9$. Similarly, the median effective radii of the metal-poor subpopulations are $R_{\rm e}=6$--$30~\kpc$, whereas the metal-rich counterparts have $R_{\rm e}=2$--$20~\kpc$. We note that increasing galaxy-to-galaxy variations towards low galaxy masses are driven by subsamples that have between $3$--$5$ GCs for which the fitting procedure performs badly, but the median values should be robust.

A similar trend of metal-poor GC systems having shallower profiles that are more extended have been observed in many observational studies \citep[e.g.][]{rhode04,bassino06,caldwell11,faifer11,pota13,kartha14,cho16,kartha16,hudson18}. It has been argued that these trends result from a scenario in which the outer metal-poor GCs formed in satellites galaxies that are accreted later on, whereas the inner metal-rich populations are mostly formed in-situ in the host galaxy \citep[e.g.][]{forbes97,brodie06}. 

\begin{figure*}
\centering
\includegraphics[width=\hsize,keepaspectratio]{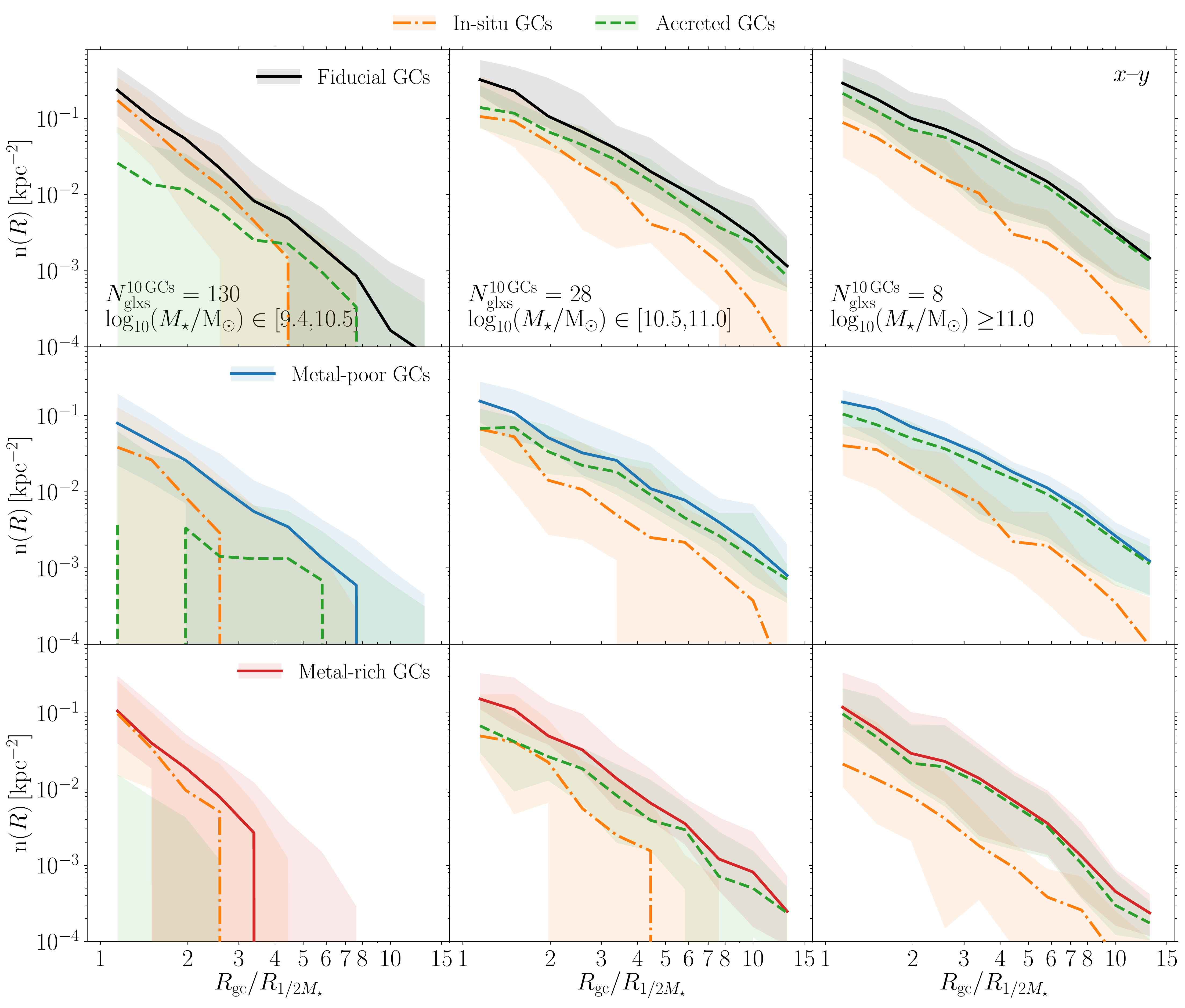}
\caption{\label{fig:r2d-feh-origin} Median projected spatial distributions of metallicity subpopulations of GCs shown as a function of their origin around the central galaxies from the \emosaics volume: fiducial populations (\textit{top row}), metal-poor GCs (\textit{middle row}) and metal-rich subpopulations (\textit{bottom row}). The spatial distributions are projected on the $x$--$y$ plane. Each column corresponds to a different galaxy stellar mass bin, and the number of galaxies within the mass bin with at least $10$ GCs ($N_{\rm glxs}^{\rm 10~GCs}$) is indicated in the bottom left side of the top row. Solid coloured lines with shaded regions correspond to medians and the $25$--$75$th percentiles of different metallicity subpopulations. For each metallicity subsample, the dash-dotted and dashed lines with shaded regions show the medians and the $25$--$75$th percentiles of the in-situ and accreted objects, respectively. Only central galaxies with at least 10 GCs within the fiducial metallicity cut are considered. Accreted objects tend to dominate the outer parts of galaxies more massive than $\log_{10}(M_{\star}/M_{\odot})\geq10.5$ (regardless of the GC metallicity), whereas lower mass galaxies host preferentially in-situ metal-rich GCs and a mixture of in-situ and accreted metal-poor GCs.}
\end{figure*}

We explore this scenario by looking at the median number density profiles of GCs for different metallicity subpopulations labeled by their origin. From top to bottom, we show in Fig.~\ref{fig:r2d-feh-origin} the median profiles of the fiducial, metal-poor and metal-rich subpopulations, respectively, in different galaxy mass bins over the sample of 166 central galaxies with at least $10~$GCs. Within each panel, we show the radial profiles of the corresponding GCs that have formed in-situ or in an accreted satellite, as well as the overall radial profile. Focusing on the fiducial GCs first, we find that low mass galaxies ($2.5\times10^{10}\leq M_{\star} \leq 3\times10^{10}~\msun$) are clearly dominated by in-situ GCs within $\sim5$--$6\times R_{1/2 M_\star}$, and they become dominated by accreted objects at larger galactocentric distances. We note that the radial profiles in this galaxy mass bin show a large scatter due to the low number of GCs hosted by these galaxies (see Fig.~\ref{fig:num-gcs-feh}). At higher galaxy masses, we find that in-situ and accreted fiducial GCs exist in comparable numbers in the inner part of the halo ($\lesssim 3\times R_{1/2M_{\star}}$), but the outskirts are dominated by the accreted GCs. The presence of large numbers of accreted GCs in the outer regions of more massive galaxies flattens the radial profiles, thus producing the observed decreasing trend of power-law slope towards brighter galaxies.

Examining the metallicity subsamples, we find the largest difference in the lowest galaxy mass bin (left column in Fig.~\ref{fig:r2d-feh-origin}). In low-mass galaxies, metal-poor GCs are a mixture of objects formed in-situ and accreted, with the latter preferentially residing in the peripheries of haloes. Contrary to that, the vast majority of metal-rich GCs form in-situ and reside in the inner region of haloes. At larger galaxy masses ($M_{\star} \geq 3\times10^{10}~\msun$), in-situ and accreted objects have comparable numbers in the inner part of the radial range considered, and the increasingly larger number of accreted GCs flattens the radial profiles. Thus, we can relate the shallower and more extended radial profiles of metal-poor GC subpopulations relative to metal-rich objects at a given galaxy mass with their preferential accreted origin. We note that the in-situ GC subpopulations also show extended radial distributions, which is likely to be caused by their early migration to an environment with a lower gas content (i.e.~galactic outskirts) where they are more likely to survive disruption until the present-day \citep{kruijssen15,keller20b}. Additionally, the increasingly larger number of accreted GCs in both metallicity subpopulations and the corresponding flattening of their radial profiles indicates that, at high galaxy masses, GCs in the halo outskirts are likely to have an accreted origin irrespective of their metallicity.

Finally, we repeat the fitting procedure on the three dimensional radial distributions of the GC metallicity subpopulations, which we show in Appendix~\ref{app:r3d}. We find that three dimensional distributions show the same trends of shallower and more extended radial profiles towards higher galaxy masses, and that there is less halo-to-halo variation than in the case of the projected profiles. 

\subsection{Dark matter}
In order to characterize the density profiles of the DM haloes of the 166 central galaxies considered, we follow the fitting procedure outlined by \citet{neto07} \citep[and also followed by][]{schaller15}. The authors suggest that the profile of the DM halo can be well characterized using a binned DM volume density profile over the radial range $r\in[0.05,1]$ times the virial radius of the halo, $r_{\rm vir}$. The fit is then performed by minimizing the root-mean-square (rms) deviation,
\be
	\sigma_{\rm fit}^2 = \dfrac{1}{N_{\rm bins}-1}\sum_{i=1}^{N_{\rm bins}} \left(\log_{10}\rho(r_{i}) - \log_{10}\rho_{{\rm DM}, i}\right)^2,
\ee 
where $N_{\rm bins} = 32$\footnote{\citet{neto07} argue that this number of bins leads to unbiased and robust results when determining the profiles of DM haloes in the Millenium Simulation, and we opt for using the same value.} corresponds to the logarithmically-spaced shells over which we calculate the DM profile, $\rho_{{\rm DM}}$. In this procedure, each radial bin is given equal weight.

We consider three functional forms to describe the volume density profile of DM, $\rho(r)$, and determine the combination of parameters that minimizes the rms deviation of each of them. The first profile that we consider is a power-law function as described above (eq.~\ref{eq:power-law}), but ensuring that the normalization is calculated in $3$D,
\be
	\rho_{\rm e} = \dfrac{3-\alpha}{4\pi (r_{\rm max}^{3-\alpha} - r_{\rm min}^{3-\alpha})}M(r_{\rm min}<r<r_{\rm max}).\label{eq:power-law-norm-3d}
\ee
Next, we consider a Navarro-Frenk-White profile \citep[NFW;][]{navarro96,navarro97} to describe our DM haloes,
\be
\rho(r) = \delta_{\rm c}\rho_{\rm cr}\left[{\dfrac{r}{r_{\rm s}} \left(1+\dfrac{r}{r_{\rm s}}\right)^2}\right]^{-1},
\ee
where $\rho_{\rm cr} = 3H^2/8\pi G = 127.5~\msun~\kpc^{-3}$ is the critical density of the Universe for closure for the cosmology used in \emosaics. The density contrast and the scale radius of the halo are given by $\delta_{\rm c}$ and $r_{\rm s}$, respectively. Lastly, we also consider that our DM haloes can be described by an Einasto profile \citep{navarro04},
\be
	\rho(r) = \rho_{\rm s} \exp\left\{-\dfrac{2}{\alpha}\left[{\left(\dfrac{r}{r_{\rm s}}\right)}^{\alpha}-1\right]\right\},
\ee
where the slope $\alpha$ changes as a function of radius. The normalization $\rho_{\rm s}$ can be calculated using the slope and the scale radius $r_{\rm s}$ as 
\be
\begin{split}
	\rho_{\rm s} =& \dfrac{\alpha e^{-2/\alpha}M(r_{\rm min}<r<r_{\rm max})}{4\pi r_{\rm s}^3 (\alpha/2)^{3/\alpha}} \times \\
 	&\left\{\gamma\left[\dfrac{3}{\alpha}, \dfrac{2}{\alpha} \left(\frac{r_{\rm max}}{r_{\rm s}}\right)^{\alpha}\right] - \gamma\left[\dfrac{3}{\alpha}, \dfrac{2}{\alpha} \left(\frac{r_{\rm min}}{r_{\rm s}}\right)^{\alpha}\right]\right\}^{-1},
\end{split}
\ee
where $\gamma(z,x)$ is the lower incomplete Gamma function. We determine the combination of free parameters that minimize the rms deviation for each radial profile: the slope $\alpha$ for the power-law function, the parameter $\delta_{\rm c}$ and the scale radius $r_{\rm s}$ for the NFW profile, and the slope $\alpha$ and the scale radius $r_{\rm s}$ for the Einasto profile. In order to ensure numerical convergence, we use the method `Trust Region Reflective' to minimize the rms deviations. This robust method is suitable for sparse problems and it allows us to place bounds on the combination of parameters explored in each profile \citep{branch99}.

Contrary to the procedure outlined by \citet{neto07} \citep[and also used by][]{schaller15}, our haloes have not been selected to be `relaxed', where relaxed haloes are sub-virial, have a low substructure mass fraction, and a small centre of mass displacement. The authors argue that these criteria are needed to avoid haloes whose density profile would not be well described by a NFW profile. However, we decide to not restrict our halo sample to avoid losing objects that could have an interesting GC population. However, this might introduce scatter in the recovered parameters, and, in some cases, result in poor fits to the DM profiles.

To estimate the quality of the fits, we calculate the rms deviation of the best-fitting profiles of the DM distributions, and we show these values in Appendix~\ref{app:fits-quality}. Out of the three functional forms considered, we find that both the NFW and the Einasto profiles provide a good description of the DM haloes, with the Einasto profile being a slightly better description. Many previous studies in the literature find similar results, and they argue that the better agreement provided by the Einasto profile is due to this profile being more accurate at describing the inner part of the DM profile compared to the NFW profile when fitting the structure of DM haloes in hydrodynamical simulations \citep[e.g.][]{navarro04,merritt06,schaller15}. 

\begin{figure}
\centering
\includegraphics[width=\hsize,keepaspectratio]{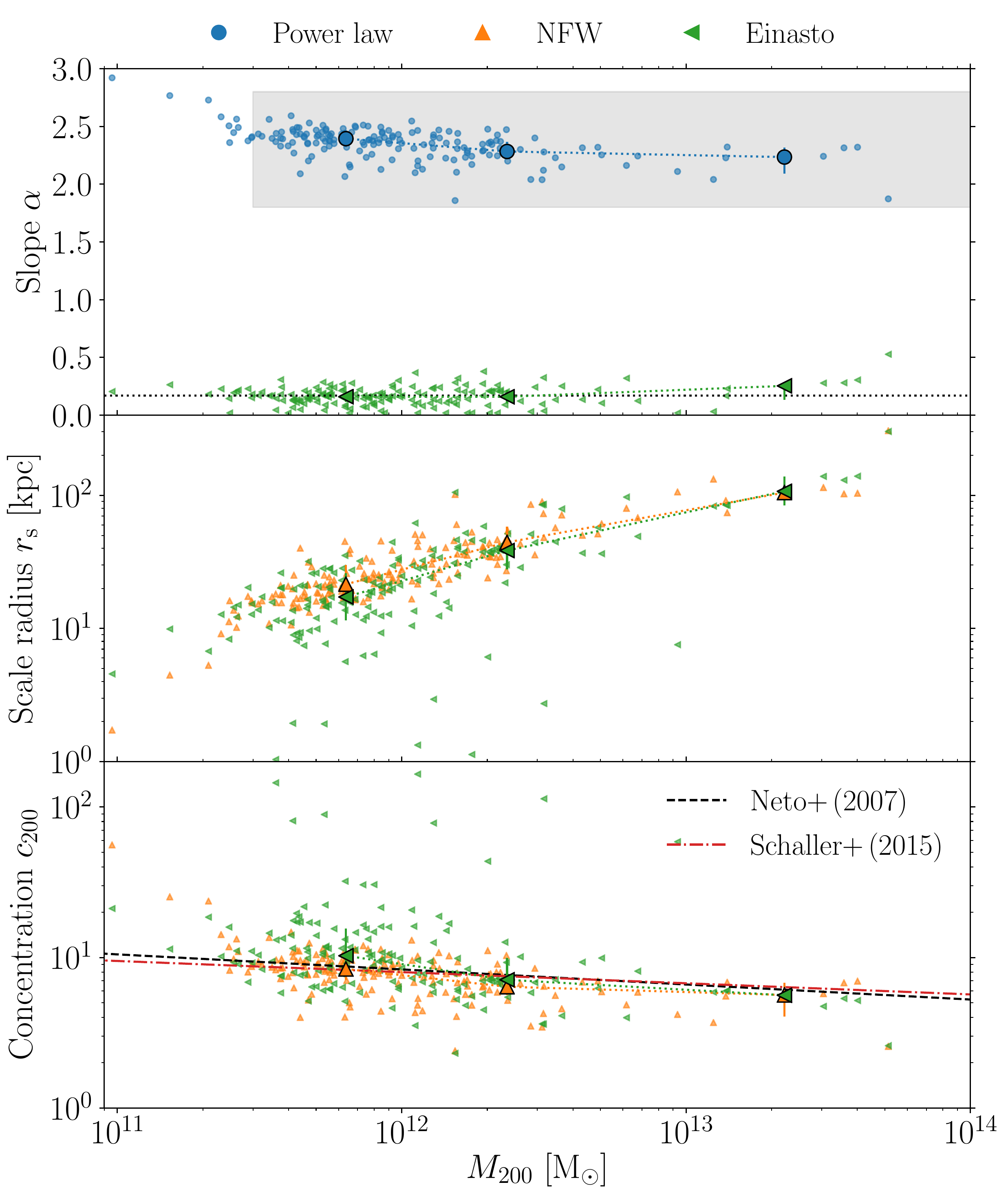}
\caption{\label{fig:fits-dm} Fitted parameters to the DM haloes of the 166 central galaxies from the \emosaics volume: slopes (\textit{top panel}), and scale radii (\textit{middle panel}) as a function of halo mass. The concentration parameters (\textit{bottom panel}) are calculated from the scale radii. Different colours indicate the parameters retrieved for each profile as indicated in the legend. The larger markers with errorbars connected by dotted lines show the median values and the $25$--$75$th percentiles. The grey-shaded region in the top panel corresponds to the power-law slopes obtained by \citet{pillepich14}, whereas the black dotted line in the same panel indicates the mass-independent Einasto slope $\alpha\approx 0.17$ \citep{navarro04}. The dashed and dash-dotted lines shown in the bottom panel correspond to fits of the concentration parameters as a function of halo mass from \citet{neto07} and \citet{schaller15}, respectively. The DM haloes of our sample of galaxies are consistent with previous literature estimates as a function of halo mass.}
\end{figure}

We show in Fig.~\ref{fig:fits-dm} the parameters of the best-fitting profiles as a function of halo mass. Focusing first on the fitted slopes of the power-law profiles, we find that they are in the range $\alpha=1.8$--$2.5$. The slopes show a mild decreasing trend towards higher halo masses, suggesting that, as discussed for the stellar and GC populations, more massive galaxies have shallower and more extended DM distributions. This is a consequence of the decrease in the mean halo concentration with increasing halo mass \citep[e.g.][]{dutton14}. We include a shaded region that corresponds to the slopes measured by \citet{pillepich14} for the \illustris DM haloes (top panel), and we find that there is excellent agreement between the slopes of our haloes and their measured values. Next, we can look at the fitted scale radii of the NFW profiles, which lie in the range $r_{\rm s}=10$--$100~\kpc$. These radii increase steeply towards higher halo masses, and show little scatter across our galaxy sample in good agreement with previous measurements from DM-only simulations \citep[e.g.][]{navarro96,navarro97}. Lastly, we examine the fitted parameters describing the Einasto profiles. Our measured slopes show very little scatter, and are consistent with the mass-independent slope $\alpha\approx 0.17$ obtained by \citet{navarro04}. In contrast, the fitted Einasto scale radii are noisier than the ones recovered from the NFW profiles, especially in haloes with $M_{200}\leq2\times10^{12}~\msun$. 

We also show in Fig.~\ref{fig:fits-dm} the concentration parameters of our DM haloes (bottom panel). We calculate these as
\be
  c_{\rm 200} = r_{200}/ r_{\rm s},
  \label{eq:c200}
\ee
using the $r_{200}$ provided by the \fof algorithm and the fitted scale radii assuming either a NFW or Einasto profile. The concentration parameters of the NFW profiles are in the range $c_{200}^{\rm NFW}=5$--$15$. Due to the noise in the scale radii, the concentration parameters from the Einasto profile range between $c_{200}^{\rm Einasto}=5$--$40$. Despite the larger scatter at low halo masses in the Einasto values, the medians for both profiles are very similar and show a mild decreasing trend with halo mass. We include the mass-concentration relations from the DM-only Millenium Simulation \citep{neto07} and from the standard resolution \eagle simulations \citep{schaller15}, which are reproduced by our median concentrated parameters. The good agreement between the best-fitting values and previous studies demonstrates that these combination of parameters and profiles provide a suitable description of our DM haloes.

\section{Tracing the structure of the DM halo}\label{sec:correlations}

In the previous section, we characterized the radial profiles of the GC populations in our simulated 166 central galaxies, as well as the shape of the DM halo profiles of their host galaxies. Now, we explore whether there are any correlations that will allow us to use number count studies of GC systems to trace the structure of the DM haloes of observed galaxies.

For this purpose, we examine correlations between the power-law slopes of the GC systems and their effective radii with the parameters describing the DM halo profiles. We focus on using the power-law slopes of the DM halo, $\alpha_{\rm PL}$, as well as the NFW profile scale radii, $r_{\rm s}^{\rm NFW}$, the extent of the halo, $r_{200}$, and the concentration parameters $c_{200}$. The structure of a DM halo is more commonly described in terms of its mass $M_{200}$ and its concentration parameter $c_{200}$, but we instead use their spatial extents, $r_{\rm s}$ and $r_{200}$, to relate with the spatial properties of GCs\footnote{These quantities can be transformed into one another using eq.~(\ref{eq:c200}) and $M_{200} = 200 \rho_{\rm cr} (4\pi/3) r_{200}^3$.}. We also looked for correlations with the parameters from the Einasto profiles, but we find these are less statistically significant, and we exclude them from the discussion.

\begin{figure*}
\centering
\includegraphics[width=\hsize,keepaspectratio]{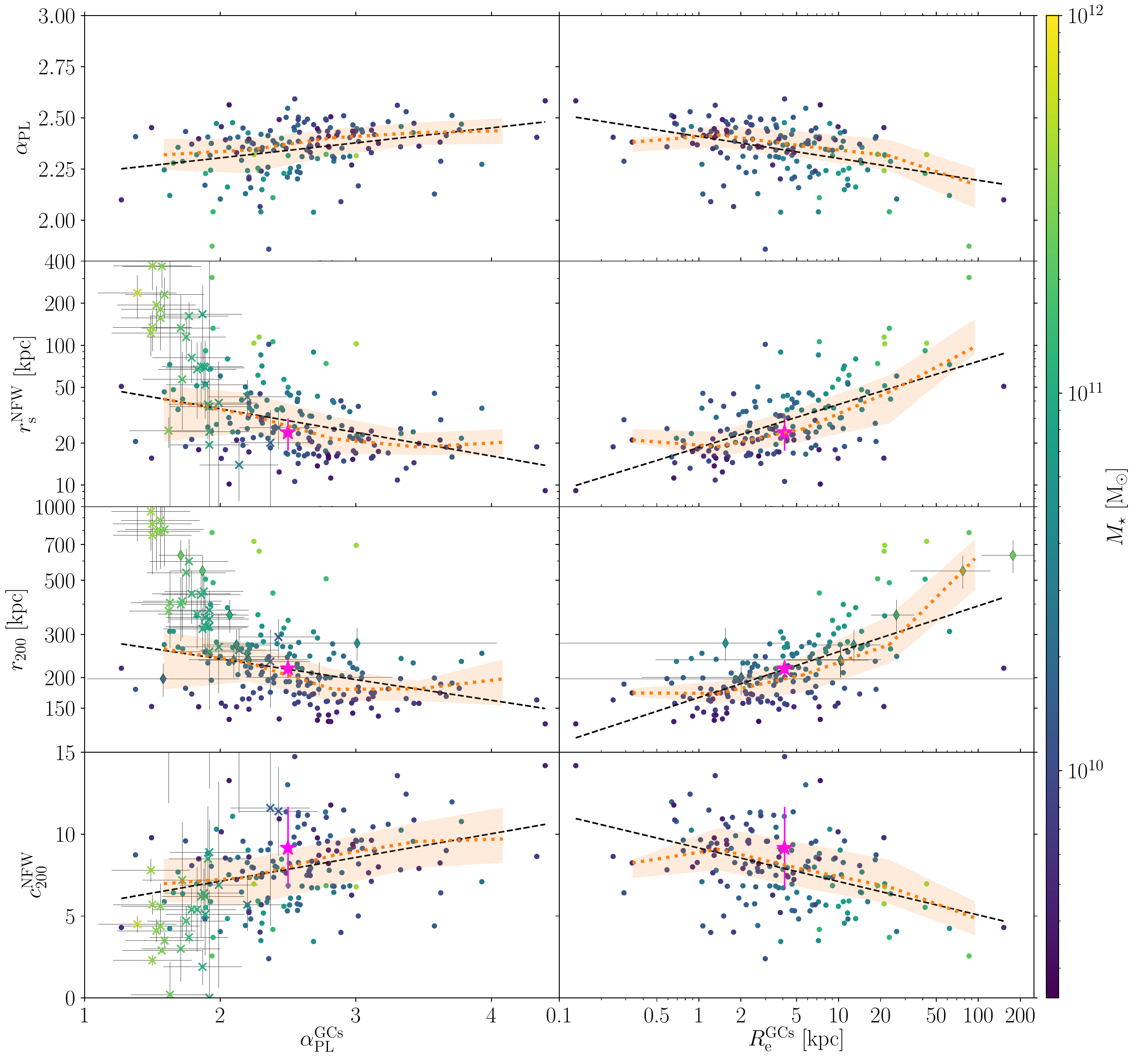}
\caption{\label{fig:corr-1dfits} Correlations between the spatial distributions of GCs and the structure of the DM haloes of their host galaxies. From top to bottom, the rows show the power-law slope, $\alpha_{\rm PL}$, the scale radii of the NFW profile, $r_{\rm s}^{\rm NFW}$, the extent of the halo, $r_{200}$, and the concentration parameter, $c_{200}$, as a function of the power-law slope and the effective radius of the fiducial GC populations. Small circles correspond to the 166 central galaxies that contain at least 10 GCs within the fiducial metallicity cut, and are colour-coded by the galaxy stellar mass. The orange dotted lines with shaded regions correspond to the median and $25$--$75$th percentiles in each panel, and the black dashed lines show the linear fits summarised in Table~\ref{tab:fits-correlations}. The Milky Way is shown by the magenta star with errorbars \citep{harris76,wolf10,hudson18,cautun20}. The crosses with errorbars correspond to the sample of ETGs from \citet{alabi16,alabi17}, and the diamonds with errorbars indicate the sample of ETGs from \citet{hudson18}. For details on the observational data, we refer the reader to the text. Both the power-law slope and the effective radius of GC populations are good tracers of the structural properties of their host DM halo.}
\end{figure*}

\begin{table*}
\centering{
  \caption{Correlations between the spatial distributions of (fiducial) GC populations and the structure of DM haloes (Figs.~\ref{fig:corr-1dfits} and~\ref{fig:corr-2dfits}). From left to right columns: independent variables included in the fits, the Spearman correlation coefficients and $p$--values, the Pearson correlation coefficients, the coefficients of the fits, the weighted rms deviation of the simulated data and of the observational samples, and the standard deviation of the simulated data, respectively. We calculate the weighted rms deviation as ${\rm WRMS} = \sqrt{(1/(N-k))\sum_{i}^N (f(\vec{x_{i}}) - z_{i})^2}$, where $N = 163$ is the number of data points, $k$ are the degrees of freedom, $z$ is the measured value and $f(\vec{x})$ is the value obtained from the fit. Using this definition, a value of zero, ${\rm WRMS}=0$, indicates a fit without scatter. The uncertainty on the properties of DM haloes inferred using GC information is thus $\sigma^2 = (N-k){\rm WRMS}^2/N$. The spatial scales are all measured in $\kpc$, and the galaxy masses are in $\msun$.}
  \label{tab:fits-correlations}
  \begin{tabular}{cccccccccc}\hline
\multicolumn{10}{c}{Linear fits: $y=ax+b$} \\ \hline
\multicolumn{2}{c}{Variables} & \multicolumn{2}{c}{Spearman} & Pearson & \multicolumn{2}{c}{Coefficients} & \multicolumn{3}{c}{Scatter} \\ \hline
$y$ & $x$ & $\rho$ & $\log_{10}(p)$ & $r_{\rm P}$ & $a$ & $b$ & WRMS & WRMS$_{\rm obs}$ & $\sigma_{y}$ \\ \hline \hline
$\alpha_{\rm PL}$ & \multirow{4}{*}{$\alpha_{\rm PL}^{\rm GCs}$}& 0.34 & -5.09 & 0.31& 0.07$\pm$0.02& 2.16$\pm$0.05& 0.12 & -- & 0.12 \\ 
$\log_{10}r_{\rm s}^{\rm NFW}$ & & -0.41 & -7.45 & -0.37& -0.17$\pm$0.03& 1.88$\pm$0.09& 0.23 & 0.75 & 0.23 \\ 
$\log_{10}r_{200}$ & & -0.38 & -6.44 & -0.31& -0.08$\pm$0.02& 2.55$\pm$0.05& 0.14 & 0.30 & 0.14 \\ 
$c_{200}$ & & 0.35 & -5.40 & 0.34& 1.4$\pm$0.3& 4.2$\pm$0.8& 2.19 & 4.56 & 2.18 \\ 
\hline
$\alpha_{\rm PL}$ & \multirow{4}{*}{$\log_{10}R_{\rm e}^{\rm GCs}$}& -0.42 & -7.60 & -0.41& -0.11$\pm$0.02& 2.41$\pm$0.01& 0.12 & -- & 0.12 \\ 
$\log_{10}r_{\rm s}^{\rm NFW}$ & & 0.57 & -15.13 & 0.60& 0.31$\pm$0.03& 1.27$\pm$0.02& 0.20 & -- & 0.20 \\ 
$\log_{10}r_{200}$ & & 0.60 & -16.92 & 0.62& 0.19$\pm$0.02& 2.22$\pm$0.01& 0.11 & 0.18 & 0.11 \\ 
$c_{200}$ & & -0.43 & -8.14 & -0.43& -2.0$\pm$0.3& 9.2$\pm$0.3& 2.11 & -- & 2.09 \\ 
\hline
\hline
\multicolumn{10}{c}{Two-parameter fits: $z=ax+by+c$} \\ \hline
\multicolumn{3}{c}{Variables} & \multicolumn{3}{c}{Coefficients} & \multicolumn{3}{c}{Scatter} & \\ \hline
$z$ & $x$ & $y$ & $a$ & $b$ & $c$ & WRMS & WRMS$_{\rm obs}$ & $\sigma_{y}$ & \\ \hline \hline
$\alpha_{\rm PL}$ & \multirow{3}{*}{$\alpha_{\rm PL}^{\rm GCs}$} & \multirow{3}{*}{$\log_{10}M_{\star}$}& 0.05$\pm$0.02& -0.11$\pm$0.02& 3.3$\pm$0.3& 0.11 & -- & 0.11 \\ 
$\log_{10}r_{\rm s}^{\rm NFW}$ & & & -0.07$\pm$0.02& 0.45$\pm$0.03& -3.0$\pm$0.3& 0.15 & 0.60 & 0.15 \\ 
$\log_{10}r_{200}$ & & & -0.012$\pm$0.006& 0.335$\pm$0.008& -1.07$\pm$0.09& 0.04 & 0.11 & 0.04 \\ 
\hline
\hline
\end{tabular}}
\end{table*}

We show in Fig.~\ref{fig:corr-1dfits} the correlations between the structural properties of the DM haloes and the power-law slopes and efective radii of the GC populations. From top to bottom, we show the correlations between the power-law slope of the DM haloes, their scale radii when assuming a NFW profile, the extent of the haloes, and their concentration parameters as a function of the power-law slope of the GC populations (left column) and their effective radii (right column). The simulated sample of 166 central galaxies is represented by the small dots and colour-coded by their galaxy stellar masses. To guide the eye, we include the median values and the $25$--$75$th percentiles in each panel, as a well as a linear fit.
Within our sample, there are three galaxies ($M_{200}\sim10^{11}~\msun$) that have lower halo masses than expected from the stellar to halo mass relation (see Fig.~\ref{fig:stellar-halo-mass}) and that have scale radii smaller by $\sim1~$dex relative to the extrapolation towards galaxies of the same mass. We find that these galaxies drive most of the scatter in the fits, and so we exclude them from Fig.~\ref{fig:corr-1dfits} and from our results. 

We summarise the statistical significance and the parameters of the linear fits in Table~\ref{tab:fits-correlations}. We find that more massive galaxies tend to have shallower DM haloes that are more extended and have lower concentration parameters, which produces shallower profiles for the GC populations with larger radial extent. The eight linear correlations explored in this figure are found to be statistically significant, with Spearman p-values below $10^{-4}$, and show relatively small scatter, except for those describing the concentration parameter. 

In order to test these correlations, we include several observational studies in Fig.~\ref{fig:corr-1dfits}. The first galaxy that we include is the Milky Way, which is represented by a magenta star with errorbars. For the structural properties of its DM halo, we use the best-fitting values derived by \citet{cautun20} in the case of a contracted DM halo. We use a projected slope for the Galactic GCs of $\alpha\simeq2.5$ \citep{harris76,wolf10}, and an effective radius of $R_{\rm e}=4.1\pm0.5~\kpc$ as determined by \citet{hudson18}. The horizontal errorbars in the right-hand column are smaller than the symbol used. The Milky Way agrees remarkably well with the median values measured in \emosaics, as well as with the correlations obtained in this work. 

The second observational sample that we show is described by \citet{alabi16,alabi17}, and is represented by the crosses with errorbars in the left-hand column of Fig.~\ref{fig:corr-1dfits}. The authors present a sample of ETGs from the SLUGGs survey for which they determine halo masses $M_{200}$ and concentration parameters $c_{200}$ using GC dynamical models, which we transform, along with their uncertainties, into the structural properties of interest here. \citet{alabi16} calculate the three-dimensional slopes of the GC populations by first deriving a relation between the de-projected slope as a function of galaxy stellar mass from a compilation of previous works (which we include in the top-right panel of Fig.~\ref{fig:fits-stars-gcs}), and then applying it to their sample of galaxies. We then transform the three-dimensional slopes for the sample of ETGs to projected slopes by substracting one dex, $\alpha_{\rm 2D} = \alpha_{\rm 3D} - 1$, and we assume the rms scatter of $0.29$ from the literature-compiled relation for their uncertainties.

The last extragalactic sample of ETGs included in Fig.~\ref{fig:corr-1dfits} is described by \citet{hudson18}, and is represented by the diamonds with errorbars. The authors obtain the halo masses $M_{200}$ from a stellar-to-halo mass relation calibrated by weak gravitational lensing \citep{hudson15}, and obtain the virial radii from $M_{200}$. The spatial properties of the GC populations are measured by fitting power-law functions and de Vaucoulers radial profiles to the total GC populations. The authors find that fitting S\'ersic profiles to their GC populations leads to noisy parameters, and decide to use instead a de Vaucouleurs profile (i.e.~a S\'ersic profile with a fixed slope of $n=4$). We find the same issue when fitting our simulated GC populations (see Sect.~\ref{sec:fit-profiles}). 

\begin{figure*}
\centering
\includegraphics[width=\hsize,keepaspectratio]{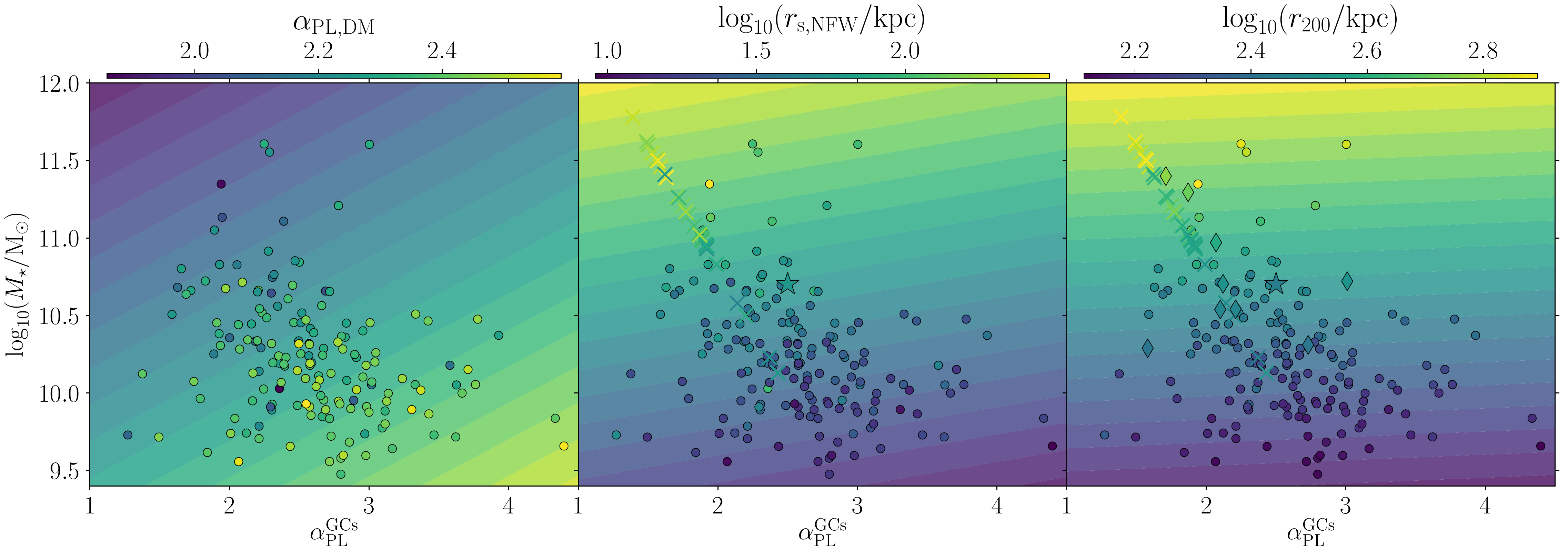}
\caption{\label{fig:corr-2dfits} Correlations between the power-law slope of fiducial GC subpopulations, the galaxy stellar mass and the structural parameters of the DM halo: power-law slopes  (\textit{left}), scale radii of the NFW profile (\textit{middle}), and the extent of DM haloes (\textit{right}). Data points are colour-coded by the properties of the DM haloes as shown by the colourbars. Small circles correspond to the sample of 166 central galaxies with at least 10 GCs within the fiducial metallicity cut from the \emosaics volume. The Milky Way is shown as a star \citep{harris76,wolf10,hudson18,cautun20}. The crosses correspond to the sample of ETGs from \citet{alabi16,alabi17}, and the diamonds indicate the sample of ETGs from \citet{hudson18}. The backgrounds correspond to the two-parameter fits performed using the simulated data and summarised in Table~\ref{tab:fits-correlations}. Including the information on the galaxy stellar mass improves the agreement with the observational sample.}
\end{figure*}

We find that the samples of ETGs follow the trends of our simulated galaxies of the same stellar mass. As these galaxies are in general more massive than the sample of central galaxies from the \emosaics volume that we consider here (see Fig.~\ref{fig:stellar-halo-mass}), their scale radii are also more extended and they are in general not encompassed within the linear fits shown in Fig.~\ref{fig:corr-1dfits} (summarised in Table~\ref{tab:fits-correlations}). The correlation between the deviation away from the fit and stellar mass that is present both in the simulated and observed galaxies, suggests that $M_{\star}$ is needed as an extra parameter. In order to overcome this caveat, we perform two-parameter fits based on the power-law slope of GCs and the galaxy stellar mass in Fig.~\ref{fig:corr-2dfits}. For these fits, we ignore the ones with the concentration parameter owing to their large scatter. 

We show in Fig.~\ref{fig:corr-2dfits} the dependence of the structural parameters of the DM haloes on the power-law slope of the fiducial GC populations and the galaxy stellar masses. Our simulated sample is represented by small circles, and the star corresponds to the observational data of the GCs in the Milky Way \citep{harris76,wolf10,cautun20}. The crosses and diamonds show the sample of ETGs from the SLUGGS survey \citep{alabi16,alabi17}, and from \citet{hudson18}, respectively. We encode the actual properties of the DM haloes in the colour of the data points. As discussed in Sect.~\ref{sec:fit-profiles}, more massive galaxies that host shallower GC radial profiles also reside in DM haloes with shallower profiles that have larger scale radii by a factor of $\sim5$--$10$. Comparing to the observational samples, we find that the sample of ETGs from the SLUGGS survey should be used with care in this parameter space due to the way that the GC spatial distributions are calculated: i.e. they use a literature-based relation of the de-projected slope of GC number density profiles with galaxy stellar mass to determine the slopes of their GC populations. This can be seen in the lack of scatter in these data in the middle and right panels of Fig.~\ref{fig:corr-2dfits}. Despite this, we find that overall the simulated galaxies follow similar trends as the observational data, suggesting that any relation obtained from the simulations can be readily applied to extragalactic observations.

Using the sample of $166$ central galaxies, we fit two-dimensional linear relations of the form $z=ax+by+c$ by least-squares minimization. We show these relations as the background colours in Fig.~\ref{fig:corr-2dfits}, and we summarise them in Table~\ref{tab:fits-correlations}. When including the galaxy stellar mass as an extra parameter, we find that the power-law slopes of the DM distributions still strongly depend on the power-law slope of the GCs distributions. The dependence on the GC spatial distribution is weaker for the scale radii, and negligible for the extent of the DM halo, such that $r_{200}$ correlates only with stellar mass. The observed stellar-to-halo mass relation shows little scatter at the mass probed by our galaxies \citep[$\sim 0.15$--$0.20~$dex, see Fig.~\ref{fig:stellar-halo-mass} and e.g.][]{hudson15}, so the galaxy stellar mass on its own is a good tracer of the extent of the DM halo. We calculate the weighted rms of these two-parameter fits, and they decrease by a factor of $\sim3$ and $\sim10$ compared to the linear fits for the scale radii and the extent of the DM halo, respectively, whereas it does not change for the power-law slope of the DM halo. This excellent agreement implies that extragalactic studies of number counts of GCs, combined with the galaxy stellar mass, can be used to trace the structure of the DM halo of their host galaxy.

\section{Conclusions}\label{sec:conclusions}

In this work, we investigate how GC number density profiles relate to the distribution of the stars and DM of their host galaxies, and whether the observed GC profiles can be used to trace the structural properties of the DM haloes of their host galaxies. For this, we use the simulated GC populations residing in a sample of 166 central galaxies from the $(34.4~\rm cMpc)^3$ periodic volume from the \emosaics project. These galaxies are required to contain at least $10$ GCs within a fiducial metallicity cut when projected onto the $x$--$y$ plane, and have masses above $M_{\star}\geq2.5\times10^9~\msun$ (see Figs.~\ref{fig:stellar-halo-mass} and \ref{fig:spatdistr}, and Table~\ref{tab:glxy-samples}).

By examining the three-dimensional spatial distributions of stars and GCs around the selected galaxies (Fig.~\ref{fig:r3d}), we find that the slope of the radial profiles becomes shallower with increasing galaxy mass. We also find that GCs are more numerous and more extended (by an order of magnitude) than satellite galaxies. This suggests that GC systems are more suitable tracers of the mass distribution in the galactic outskirts than satellite galaxies. We then project the stellar and GC populations in our galaxies along three different axes, i.e.~face-on, random and edge-on, and study the sphericity of their projected radial profiles (Fig.~\ref{fig:r2d-ratios}). We find that the populations of stars and GCs are less spherical in less massive galaxies ($2.5\times10^{10}\leq M_{\star}\leq 3\times10^{10}~\msun$), with GC populations showing signs of slight prolate distributions. Given the low number of GCs hosted in these galactic systems (see Fig.~\ref{fig:num-gcs-feh}), their spatial distributions are not fully sampled, and thus properly modelled using spherical distributions. 

We then study the projected spatial distributions of stars and GCs assuming random galaxy orientations for different metallicity subpopulations (Fig.~\ref{fig:r2d-feh}). While the metal-rich stellar populations are found to dominate the radial profile within the radial range considered, the GC subpopulations show a metallicity gradient across the galaxy mass range probed by our sample of galaxies. The metal-rich GC subpopulations dominate the inner parts of galaxies, while metal-poor objects become more numerous in the outer regions. Similar metallicity gradients have long been observed in extragalactic GC systems \citep[e.g.][]{rhode04,bassino06,caldwell11,faifer11,pota13,kartha14,cho16,kartha16,hudson18}, in good agreement with our results. 

We quantify the projected spatial profiles of stars and GCs by fitting power-law distributions, as well as de Vaucouleurs profiles, using a maximum likelihood formalism. We apply this fitting procedure to the fiducial samples around the 166 central galaxies, and repeat it for different metallicity subpopulations (Fig.~\ref{fig:fits-stars-gcs}). We find that more massive galaxies host stellar and GC populations with shallower radial profiles that also have larger effective radii. Similar trends are found for the different metallicity subpopulations across our galaxy mass range. We also find that metal-poor subpopulations have, on average, shallower and more extended profiles than metal-rich GCs. The increasing galaxy-to-galaxy variation in the properties of GC populations towards low galaxy stellar masses is due to the low number of objects (see Appendix~\ref{app:fits-quality}). 

We then explore whether these trends are due to the assembly history of the GC populations. For that, we examine the projected number density profiles of GCs for different metallicity subpopulations of different origin (i.e.~in-situ or accreted, Fig.~\ref{fig:r2d-feh-origin}). We find that the shallower slopes and more extended profiles with more massive galaxies is the result of these galaxies assembling their halo via the accretion of satellite galaxies \citep[e.g.][]{qu17} that preferentially deposit their GC populations in the outskirts. We suggest that the larger extent of the metal-poor GCs is due to two reasons. First, the metal-poor GC subpopulations tend to have a predominantly accreted origin across our galaxy mass range, and so the metal-poor objects in the peripheries are mostly of accreted origin. In massive galaxies, the larger extent of the in-situ metal-poor GCs is related to their survival, i.e.~only those that migrate towards the gas-poor environments of the outer regions can survive to the present day. In contrast, metal-rich subpopulations in low mass galaxies are predominantly in-situ, but they also become dominated by accreted objects towards more massive galaxies, thus producing a flattening of their radial profiles. Thus, we note that the trend of increasing effective radius with galaxy mass is found in all metallicity subpopulations, implying that metallicity alone does not indicate an accretion origin of a given object. 

We describe the DM haloes within which our sample of 166 central galaxies reside with power-law density profiles, Navarro-Frenk-White profiles and Einasto profiles (Fig.~\ref{fig:fits-dm}). As seen in previous studies, we find that our simulated DM haloes have shallower profiles and larger extents as their masses increase \citep[e.g.][]{navarro96,navarro97,pillepich14}. The concentration parameters of the DM haloes show a shallow decreasing trend with their mass, in good agreement with previous results from both DM only and hydrodynamic simulations \citep[e.g.][]{neto07,schaller15}.

Finally, we study whether the spatial distributions of GCs trace the structural properties of the DM haloes of their host galaxies. For this, we explore relations between the power-law slopes of the DM haloes, the scale radii when assuming a NFW profile, the extent of the DM halo and the concentration parameters and the power-law slopes and effective radii of the fiducial GC populations (Fig.~\ref{fig:corr-1dfits}). We summarise the one-dimensional fits obtained in Table~\ref{tab:fits-correlations}. We find that both the power-law slopes of GCs and their effective radii are good predictors of the structure of the DM halo.

We compare the simulations to observational samples of GC systems in ETGs in Fig.~\ref{fig:corr-1dfits} \citep{alabi16,alabi17,hudson18}, as well as for the GCs in the Milky Way \citep{harris76,wolf10,hudson18,cautun20}. We find that the observational samples follow the same trends as the simulated galaxies of similar mass. Previous observational studies have found similar trends in massive ETGs \citep[e.g.][]{kartha14,hudson18,forbes18}, and in this study we extend the analysis towards lower galaxy stellar masses. 

The one-parameter fits obtained do not fully capture the behaviour at the high galaxy mass end, so we test the effect of including stellar mass as a second parameter in the fits (Fig.~\ref{fig:corr-2dfits}). Including the galaxy mass shows that stellar mass is a better predictor of the halo virial extent, and reduces the scatter in our fits by a factor of $\sim 3$ and $\sim9$ in the relations for the scale radii and the extent of the DM halo, respectively. These two-dimensional fits are also provided in Table~\ref{tab:fits-correlations}.

The good agreement with the observational samples suggests that we can use the projected number counts of GC populations, alongside their galaxy stellar masses, to trace the structure of their host DM haloes in the Local Universe. This result is highly promising as mapping bright GC populations out to large galactocentric distances is much less observationally demanding than observing the faint and diffuse stellar halo in the galactic outskirts. Additionally, galaxies host GC populations that tend to be an order of magnitude more numerous than their satellite galaxies \citep[e.g.][]{geha17,mao21}, thus making GCs ideal probes of the outer matter distribution and the DM halo of their host galaxy.

\section*{Acknowledgements}
MRC gratefully acknowledges the Canadian Institute for Theoretical Astrophysics (CITA) National Fellowship for partial support. MRC also gratefully acknowledges a Fellowship from the International Max Planck Research School for Astronomy and Cosmic Physics at the University of Heidelberg (IMPRS-HD). MRC, STG and JMDK gratefully acknowledge funding from the European Research Council (ERC) under the European Union’s Horizon 2020 research and innovation programme via the ERC Starting Grant MUSTANG (grant agreement number 714907).
AD is supported by a Royal Society University Research Fellowship. AD also acknowledges support from the Leverhulme Trust and the Science and Technology Facilities Council (STFC) [grant numbers ST/P000541/1, ST/T000244/1]. 
JMDK gratefully acknowledges funding from the German Research Foundation (DFG) in the form of an Emmy Noether Research Group (grant number KR4801/1-1). 
JLP is supported by the Australian government through the Australian Research Council’s Discovery Projects funding scheme (DP200102574).
RAC is a Royal Society University Research Fellow.
NB gratefully acknowledges financial support from the European Research Council (ERC-CoG-646928, Multi-Pop) as well as from from the Royal Society (University Research Fellowship).

This work used the DiRAC Data Centric system at Durham University, operated by the Institute for Computational Cosmology on behalf of the STFC DiRAC HPC Facility (www.dirac.ac.uk). This equipment was funded by BIS National E-infrastructure capital grant ST/K00042X/1, STFC capital grants ST/H008519/1 and ST/K00087X/1, STFC DiRAC Operations grant ST/K003267/1 and Durham University. DiRAC is part of the National E-Infrastructure. The work also made use of high performance computing facilities at Liverpool John Moores University, partly funded by the Royal Society and LJMU’s Faculty of Engineering and Technology.

\textit{Software}: This work made use of the following \textsc{Python} packages: \textsc{h5py} \citep{h5py_allversions}, \textsc{Numpy} \citep{vanderWalt11}, \textsc{Pandas} \citep{ mckinney-proc-scipy-2010, pandas_allversions}, \textsc{Pynbody} \citep{pynbody} and \textsc{Scipy} \citep{jones01}, and all figures have been produced with the library \textsc{Matplotlib} \citep{hunter07}. 

%%%%%%%%%%%%%%%%%%%%%%%%%%%%%%%%%%%%%%%%%%%%%%%%%%
\section*{Data Availability}

The data underlying this article will be shared on reasonable request to the corresponding author.



\bibliographystyle{mnras}
\interlinepenalty=10000 % avoid breaking hyperref links over two pages
\bibliography{bibdesk-bib}

%%%%%%%%%%%%%%%%%%%%%%%%%%%%%%%%%%%%%%%%%%%%%%%%%%

%%%%%%%%%%%%%%%%% APPENDICES %%%%%%%%%%%%%%%%%%%%%

\appendix

\section{Quality of the fitting procedure}\label{app:fits-quality}

In this appendix, we quantify the quality of the fitting procedure performed in Sect.~\ref{sec:fit-profiles} to the spatial distributions of GCs, stars and DM. 

\begin{figure}
\centering
\includegraphics[width=\hsize,keepaspectratio]{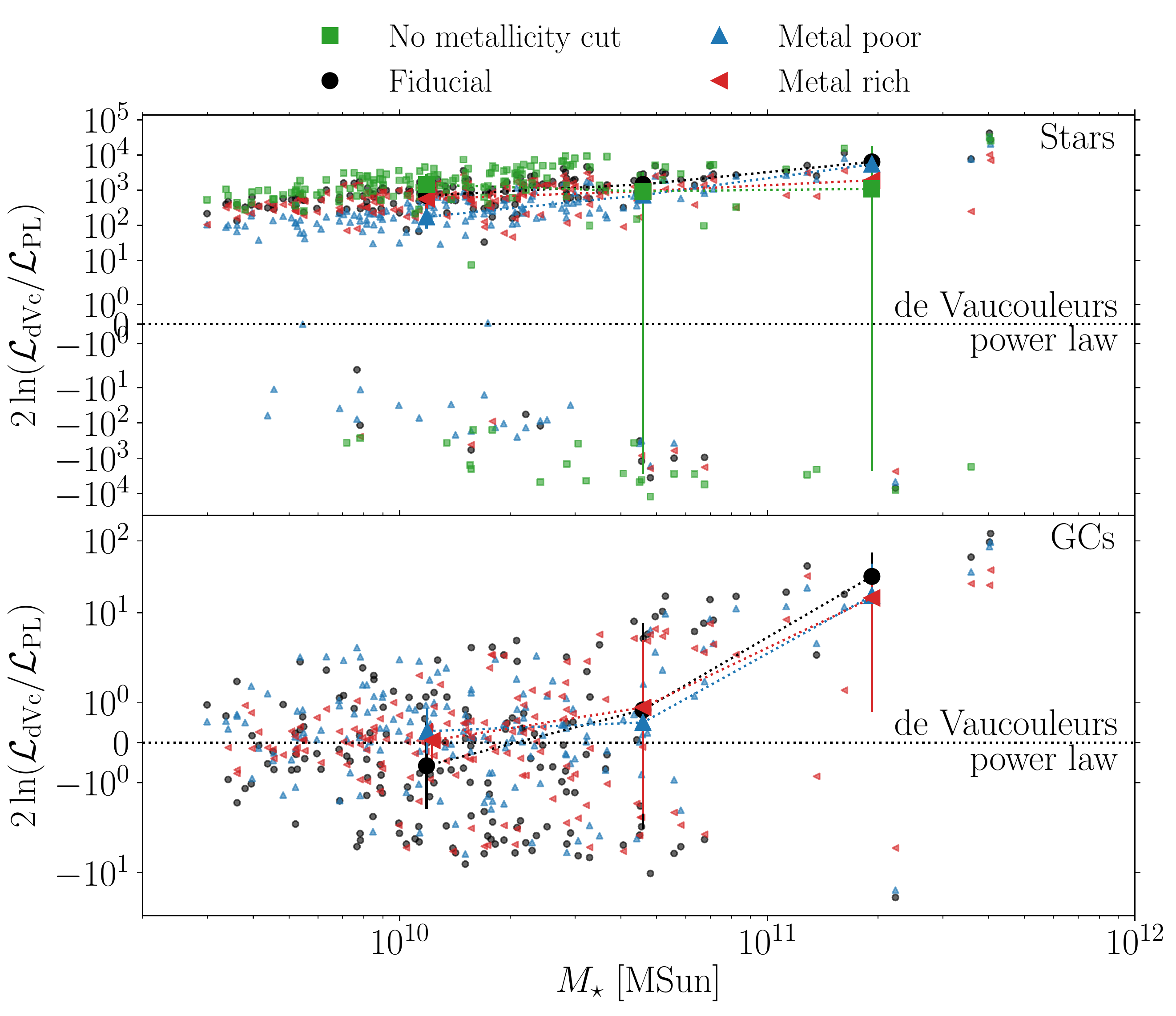}
\caption{\label{fig:app-r2d-fits-gcs-llmin} Difference between the log-likelihoods of the best-fitting de Vaucouleurs and power-law profiles to the projected radial distributions of stars (\textit{top panel}) and GCs (\textit{bottom panel}) in different metallicity cuts around central galaxies from the \emosaics volume. Metallicity subpopulations are indicated by different small coloured markers as stated in the legend. Big markers with errorbars connected by dotted lines show the median values and $25$--$75$th percentiles.}
\end{figure}

Here, we calculate the log-likelihood of the best-fitting power-law and de Vaucouleurs profiles for each metallicity subpopulation of stars and GCs. We show the difference between the log-likelihoods in Fig.~\ref{fig:app-r2d-fits-gcs-llmin}. The performance of our minimimum log-likelihood fitting procedure improves for more massive galaxies that contain a larger number of particles, as these sample the entire radial range considered. More sparsely populated subpopulations, such as metal-poor GCs in lower mass galaxies, tend to produce lower quality fits than the more metal-rich subpopulations. The stellar populations are better described by the de Vaucouleurs radial profiles across galaxy stellar mass. Contrary to that, power-law functions are a better description of the GCs subpopulations in lower mass galaxies, up to $M_{\star}\simeq4\times10^{10}~\msun$ for the metal-poor GC systems.

\begin{figure}
\centering
\includegraphics[width=\hsize,keepaspectratio]{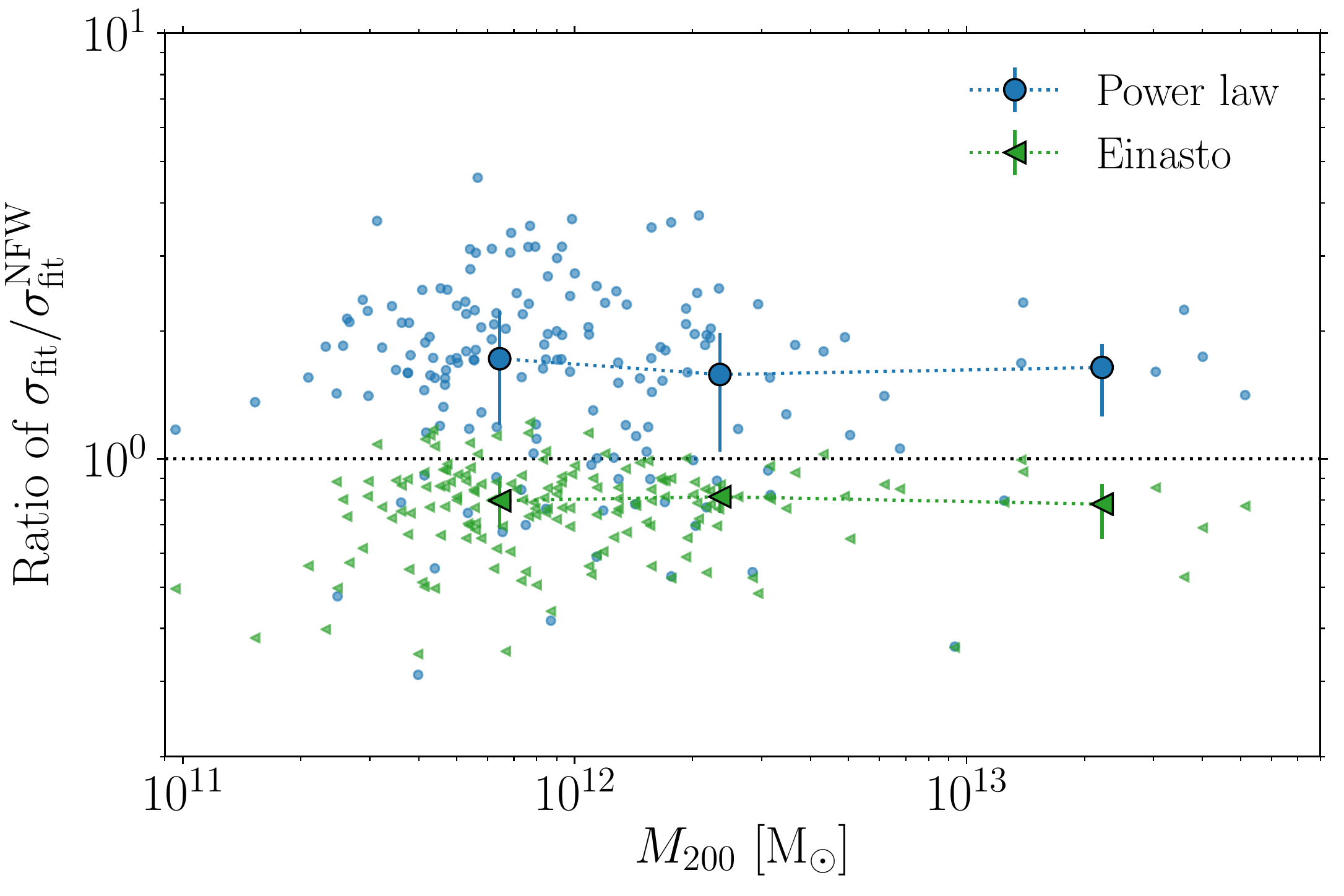}
\caption{\label{fig:app-r2d-fits-dm-rmsfit} Ratio of the rms of the best-fitting DM halo profiles relative to the NFW profile as a function of the mass of the halo. Data points show the fits performed to the DM haloes surrounding the 166 central galaxies from the \emosaics volume. Big markers with errorbars connected by dotted lines show the median values and $25$--$75$th percentiles.}
\end{figure}

We also quantify the quality of the fits performed to the DM haloes by calculating the rms deviation of each of the best-fitting profiles considered. We show in Fig.~\ref{fig:app-r2d-fits-dm-rmsfit} the ratio of the rms of the power-law and Einasto fits over the NFW fits. Since our fitting procedure is based on the binned DM profiles, there is no trend between the ratio of the rms fits with increasing galaxy mass.

Out of the three functional forms considered, we find that both the NFW and the Einasto profiles provide a good description of our haloes, with the Einasto profile being a slightly better description. Many previous studies in the literature find similar results, and they argue that the better agreement provided by the Einasto profile is due to this profile being more accurate at describing the inner part of the DM profile compared to the NFW profile \citep[e.g.][]{navarro04,merritt06,schaller15}. 

\section{Different radial ranges}\label{app:diff-radial-range}

\begin{figure}
\centering
\includegraphics[width=\hsize,keepaspectratio]{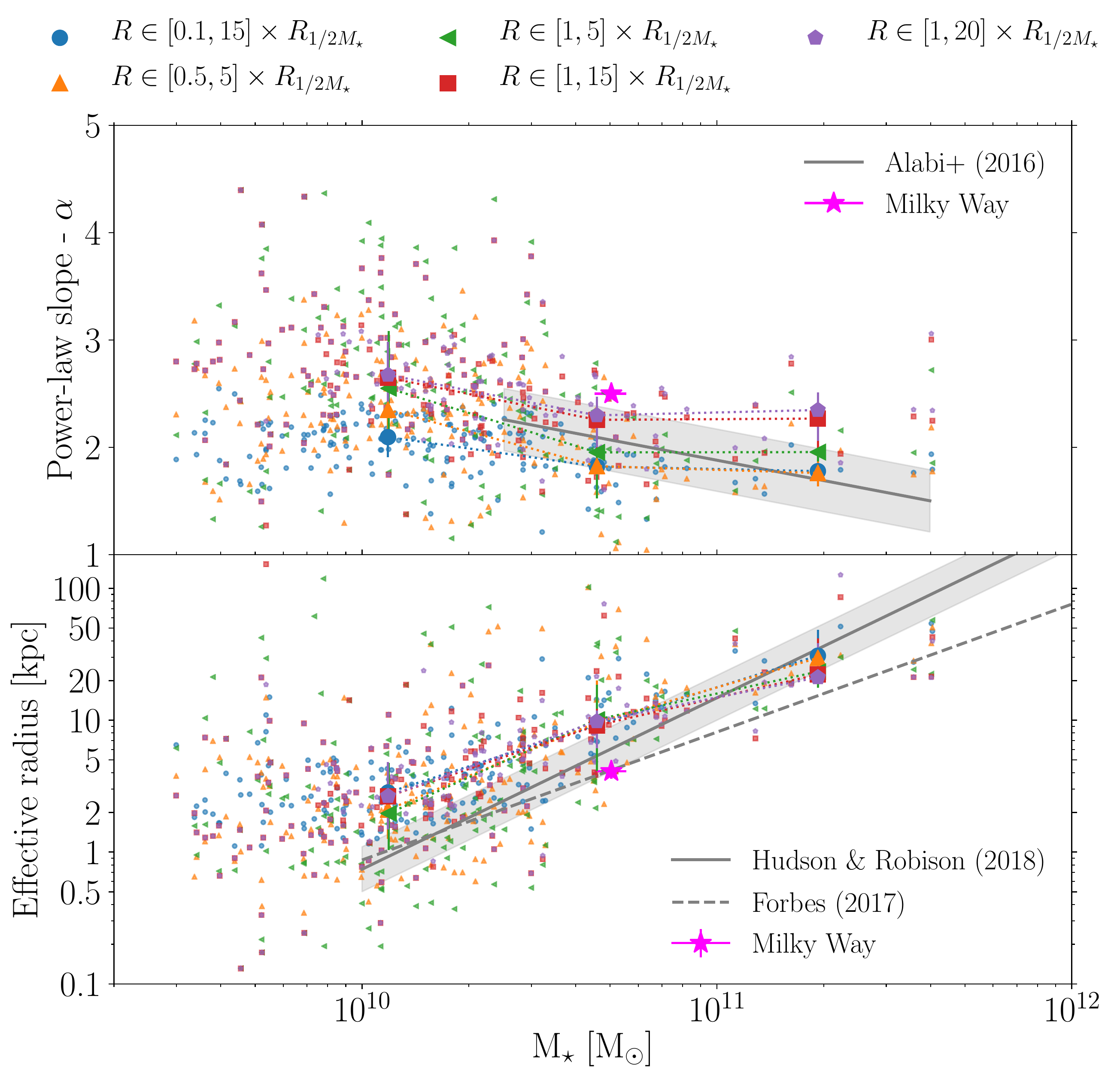}
\caption{\label{fig:app-radrange} Characterizing the projected number density radial profiles of the fiducial GC populations over different radial ranges: power-law slope (\textit{top panel}) and effective radius (\textit{bottom panel}) as a function of galaxy stellar mass. Data points of different colours correspond to the radial ranges indicated in the legend, and the big markers with errorbars connected by dotted lines show the median values and $25$--$75$th percentiles. The magenta star corresponds to the GCs in the Milky Way \citep{harris76,wolf10,hudson18,cautun20}, and the solid line with a shaded region in the top panel corresponds to the fit described by \citet{alabi16}. The solid and dashed grey lines in the bottom panel correspond to the fits obtained by \citet{hudson18} and \citet{forbes17}, respectively.}
\end{figure}

Here, we repeat the fitting procedure outlined in Sect.~\ref{subsec:calc-rad-prof} to characterize the radial distributions of the fiducial GC systems over different radial ranges. We show in Fig.~\ref{fig:app-radrange} the fitted power-law slopes and effective radii of GC populations as a function of galaxy stellar mass. We find that the slopes of the power-law profiles change as a function of the radial range because the true distribution flattens in the center of the galaxy. Including the inner half-mass stellar radius of the galaxy leads to shallower radial profiles, and extending the fit to the outer part of the GC populations steepens the radial profiles across all galaxy masses. This suggests that simple power-law distributions are not the best description of the radial profile of GCs when the inner part of the galaxy is probed, and that more complex distributions such as a S\'ersic profile \citep{sersic63,sersic68} should be considered \citep[e.g.][]{faifer11,pota13}. 

Contrary to the slope, the effective radii of the GC populations are quite insensitive to the choice of the radial range. We only find that small radial ranges (e.g.~between $1$--$5\times R_{1/2M_{\star}}$) show an increased amount of scatter because the de Vaucouleurs profiles are less adequate when applied only to the inner halo. Given that the underdisrupted GC populations are expected to reside preferentially in the inner part of the galaxy, and that observational studies tend to avoid that region due to crowding, we maintain the inner limit at one stellar half-mass radius in our main analysis. 

\section{Three dimensional distributions}\label{app:r3d}

We repeat in this appendix the analysis performed in Sect.~\ref{sec:fit-profiles} to characterize the three-dimensional number density profiles of GCs using power-law and de Vaucouleurs functions. We modify appropriately the normalization of the power-law functions (eq.~\ref{eq:power-law-norm-3d}) and of the de Vaucouleurs profiles, 
\be
\begin{split}
  f_{\rm e} =& \dfrac{b_{4}^{12} e^{-b_4}}{12\pi r_{\rm e}^3 } \times \\
  &\left\{\gamma\left[12, b_{4} \left(\frac{r_{\rm max}}{r_{\rm e}}\right)^{\frac{1}{4}}\right] - \gamma\left[12, b_{4} \left(\frac{r_{\rm min}}{r_{\rm e}}\right)^{\frac{1}{4}}\right]\right\}^{-1},
\end{split}
\ee
as well as the probability of finding an object $i$ at a radius $r_{i}$ given a profile $f(r)$, $\mathcal{P}(r_{i})$,
\be 
\mathcal{P} (r_{i}) = 4\pi r_{i}^{2} f(r_{i}).
\ee
 
\begin{figure}
\centering
\includegraphics[width=\hsize,keepaspectratio]{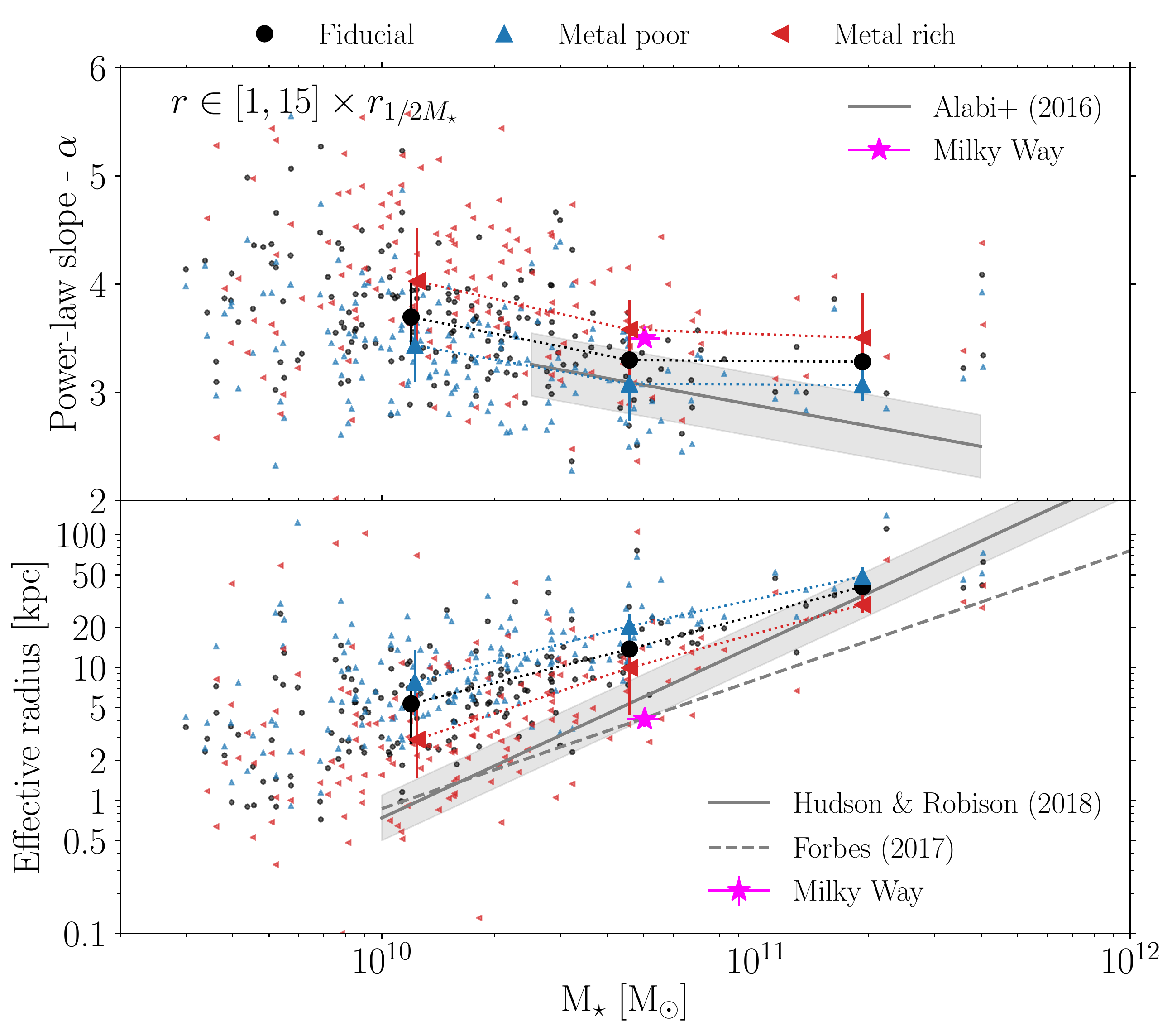}
\caption{\label{fig:app-r3d-fits-gcs} Three dimensional radial profiles of different metallicity subpopulations of GCs around central galaxies from the \emosaics volume: power-law slope of the number density radial profile of GCs (\textit{top panel}), and effective radius of each subpopulation (\textit{bottom panel}), as a function of galaxy stellar mass. Data points correspond to the 166 central galaxies with at least 10 GCs within the fiducial metallicity cut. Metallicity subpopulations are indicated by different small coloured markers. Big markers with errorbars connected by dotted lines show median values and $25$--$75$th percentiles. The magenta star corresponds to the GCs in the Milky Way \citep{harris76,wolf10,hudson18,cautun20}, and the solid line with a shaded region in the top panel corresponds to the observational relation to de-projected slopes described by \citet{alabi16}. The solid and dashed grey lines in the bottom panel correspond to the observational fits obtained by \citet{hudson18} and \citet{forbes17}, respectively.}
\end{figure}

We show in Fig.~\ref{fig:app-r3d-fits-gcs} the fitted power-law slopes $\alpha$ for the GC metallicity subpopulations as a function of galaxy stellar mass, as well as their effective radii. As in the case of the projected distributions discussed in the main body of this article, we find that more massive galaxies present shallower GC distributions that are more extended. For a given galaxy stellar mass, we find again that metal-poor subpopulations have shallower and more extended profiles than the metal-rich subpopulations. In contrast to the projected profiles, we find that the three-dimensional distributions show less scatter in the low mass regime ($M_{\star}\lesssim4\times 10^{10}~\msun$). 

%%%%%%%%%%%%%%%%%%%%%%%%%%%%%%%%%%%%%%%%%%%%%%%%%%

% Don't change these lines
\bsp	% typesetting comment
\label{lastpage}
\end{document}